\documentclass[11pt,tightenlines,eqsecnum,floats,aps,amsmath,amssymb,nofootinbib,prd,shownopacs,floatfix]{revtex4}

\usepackage{graphicx}
\usepackage{epstopdf}
\usepackage{latexsym}
\usepackage{amssymb}
\usepackage{amsmath}
\usepackage{color}
\usepackage{mathrsfs}
\usepackage{xparse}
\usepackage{float}
\usepackage{mathtools}

\usepackage[center]{subfigure}

\begin{document}
  \renewcommand\arraystretch{2}
 \newcommand{\bq}{\begin{equation}}
 \newcommand{\eq}{\end{equation}}
 \newcommand{\bqn}{\begin{eqnarray}}
 \newcommand{\eqn}{\end{eqnarray}}
 \newcommand{\nb}{\nonumber}
 \newcommand{\lb}{\label}
 \newcommand{\cb}{\color{blue}}
    \newcommand{\cc}{\color{cyan}}
        \newcommand{\cm}{\color{magenta}}
\newcommand{\rc}{\rho^{\scriptscriptstyle{\mathrm{I}}}_c}
\newcommand{\rd}{\rho^{\scriptscriptstyle{\mathrm{II}}}_c} 
\NewDocumentCommand{\evalat}{sO{\big}mm}{%
  \IfBooleanTF{#1}
   {\mleft. #3 \mright|_{#4}}
   {#3#2|_{#4}}%
}
\newcommand{\PRL}{Phys. Rev. Lett.}
\newcommand{\PL}{Phys. Lett.}
\newcommand{\PR}{Phys. Rev.}
\newcommand{\CQG}{Class. Quantum Grav.}
\newcommand{\parallelsum}{\mathbin{\!/\mkern-5mu/\!}}

\title{Primordial power spectrum from a matter-Ekpyrotic bounce scenario in loop quantum cosmology}
\author{Bao-Fei Li $^{1}$}
\email{baofeili1@lsu.edu}
\author{Sahil Saini $^{2}$}
\email{sahilsaiini@gjust.org}
\author{Parampreet Singh$^1$}
\email{psingh@lsu.edu}
\affiliation{
$^1$ Department of Physics and Astronomy, Louisiana State University, Baton Rouge, LA 70803, USA\\
$^2$ Department of Physics, Guru Jambheshwar University of Science $\&$ Technology, Hisar 125001, Haryana, India
}

\begin{abstract}
A union of matter bounce and Ekpyrotic scenarios is often studied in an attempt to combine the most promising features of these two models. Since non-perturbative quantum geometric effects in loop quantum cosmology (LQC) result in natural bouncing scenarios without any violation of energy conditions or fine tuning, an investigation of matter-Ekpyrotic bounce scenario is interesting to explore in this quantum gravitational setting. In this work, we  explore this unified phenomenological model  for  a spatially flat Friedmann-Lema\^itre-Robertson-Walker (FLRW) universe in LQC filled with  dust  and a scalar field in an Ekpyrotic scenario like  negative potential. Background dynamics and  the power spectrum of the comoving curvature perturbations are numerically analyzed with various initial conditions and a suitable choice of the initial states. By varying the initial conditions we consider different cases of dust and Ekpyrotic field domination in the  contracting phase.   We use the dressed metric approach to numerically compute the primordial power spectrum of the comoving curvature perturbations which turns out to be almost scale invariant for the modes which exit the horizon in the matter-dominated phase. But, in contrast with a constant magnitude power spectrum obtained under approximation of a constant Ekpyrotic equation of state using deformed algebra approach in an earlier work, we find that the magnitude of power spectrum changes during evolution. Our analysis shows that  the bouncing regime only leaves imprints on the modes  outside the scale-invariant regime.  However, an analysis of the spectral index  shows inconsistency with the observational data, thus making further improvements in such a  model necessary.
\end{abstract}
\maketitle
\section{Introduction}
\renewcommand{\theequation}{1.\arabic{equation}}\setcounter{equation}{0}
While the inflationary scenario provides an excellent description of the early universe including the generation of the scale invariant spectrum of perturbations, it is past incomplete because of the big bang singularity. Different models have been proposed based on bouncing cosmologies which attempt to provide a viable non-singular description of the very early universe. Due to the duality relation for the generation of scale invariant spectrum of perturbations between an inflationary epoch in the expanding branch and the dust (matter) dominated epoch in the contracting branch \cite{Wands1999}, the matter bounce scenario has been advocated as an alternative to inflation where a scale invariant power spectrum is produced by curvature perturbations that exit the horizon during the matter dominated contraction phase \cite{FinelliBB2002,Brandenberger2009,CaiSaridakis2011}. 
But, a problematic feature of the matter bounce scenario is the Belinski-Khalatnikov-Lifshitz (BKL) instability during the contracting phase \cite{BKL1970}. The BKL instability occurs due to growth of anisotropies during the contracting phase which come to dominate the dynamics of the universe near the classical singularity and play an important role in chaotic Mixmaster behavior. Small departures from perfect isotropy or anisotropic quantum fluctuations in the contracting phase may lead to the BKL instability unless the initial conditions are fine-tuned to avoid such a scenario. 

The BKL instability is avoidable if the dynamics near the classical singularity in the contracting branch is dictated by a fluid which can overcome the growth of anisotropies. Recall that the anisotropic shear scales as $a^{-6}$, where $a$ denotes the scale factor, and thus it grows faster than the energy density of all fluids except those with stiff and ultra-stiff equations of state. But even in the case of stiff matter in presence of anisotropies one requires a considerable amount of fine tuning to obtain a point like or an isotropic approach to singularity in the classical theory \cite{jacobs} or even bounce in a quantum gravity framework \cite{kasner}. Thus, to obtain an effective  isotropization it is important to include a fluid with equation of state greater than unity which dictates the dynamics in the contracting branch for a sufficiently long time. Such a fluid description arises naturally in the presence of negative potentials such as in Ekpyrotic scenarios \cite{KOSteinTurok2001,KOSSteinTurok2002}. This scenario originally motivated by the inter-brane dynamics in a higher dimensional bulk has a negative exponential potential. As the branes approach a collision, the behavior of the moduli field in the negative potential is such that the equation of state can become much greater than unity. 
Thus the Ekpyrotic field can dominate over anisotropies at small scale factors in the contracting phase, and  avoid the BKL instability \cite{TurokSteinhardt2004} (see \cite{Bozza2009} for an alternative way to avoid the BKL instability)\footnote{This may not hold in the presence of anistropic pressures \cite{BarrowYamamoto2010}.}. This interesting result has been replicated in more general Ekpyrotic scenarios which are not necessarily motivated by bulk-brane settings and have a different form of the negative potential \cite{CaiBBPP2013}.
But, a problematic feature of the Ekpyrotic scenario is that a universe inhabited by a single Ekpyrotic field alone cannot produce a scale invariant spectrum \cite{Lyth2002,FinelliBB2002b,Hwang2002,MartinPPSchwarz2002}. Further, the non-singular transition is often assumed in this model whose realization along with a required turn around of the moduli field in the potential to start a new cycle is difficult even in presence of quantum gravitational effects \cite{PSKV1}. To overcome the issue of big crunch singularity,  ``new Ekpyrotic scenario" with two Ekpyrotic fields has been proposed which can produce a scale invariant spectrum \cite{BKO2007,Creminelli2007}. However, the new Ekpyrotic scenario has been shown to suffer from the same anisotropic instability problem discussed above, along with a blue spectrum resulting from an adiabatic mode which spoils scale invariance  \cite{XueSteinhardt2010,XueSteinhardt2011}.

Given that the matter bounce and Ekpyrotic scenarios can solve different problems in the bouncing cosmologies, their union has often been considered \cite{CaiBB2012,haro2}. In the simplest setting of such a construction, 
a scale invariant spectrum can be obtained in the dust dominated epoch while the Ekpyrotic phase alleviates the problem of anisotropies near the bounce. However, a crucial and non-trivial ingredient in any such combined setting, or the matter bounce and Ekpyrotic scenarios by themselves is the occurrence of a non-singular bounce which allows the scale invariant power spectrum to pass to the expanding branch without substantially changing its character. As in Ekpyrotic models, a  variety of avenues have been explored for realizing such a bounce in matter bounce scenarios. These include using Horava-Lifshitz gravity \cite{Brandenberger2009}, $f(T)$ gravity \cite{CaiSaridakis2011}, or models that rely on violating the null energy condition near the bounce such as ghost condensate bounce \cite{LinBB2010}, quintom bounce \cite{CaiZhang2009} or Galilean bounce \cite{Lehners2014}. But, most of these studies so far exclude the quantum gravity effects which are expected to play an important role in the resolution of cosmological singularities.  These expectations turn out to be true in the framework of loop quantum cosmology (LQC) \cite{ASstatusreport}, where a bounce occurs due to non-perturbative quantum gravity effects 
\cite{APS,slqc} and a non-singular evolution is obtained via a quantum difference equation even in the presence of non-trivial potentials \cite{gls2020}. Incidentally, LQC allows an effective spacetime description 
which has been shown to be an excellent approximation to the underlying quantum gravitational dynamics \cite{APS,DGS2014a,DGS2014b,DJMS2017,ps18}. Using this effective dynamics one finds that cosmological singularities are generically resolved for isotropic and anisotropic spacetimes for all perfect fluids without any violation of null energy condition \cite{generic}.

It is therefore natural to understand the matter bounce and Ekpyrotic scenarios in the framework of LQC. Using effective dynamics of LQC, a non-singular model with an Ekpyrotic potential was obtained \cite{bms,PSKV1} but it was found that a viable model where the moduli field turns around in the negative potential cannot be realized unless another matter field or anisotropies are present \cite{PSKV2}. Investigations concerning a fluid with a fixed equation of state which is ultra-stiff have also been made \cite{WE2}. The matter bounce scenario was studied in LQC earlier \cite{WE1} and it was found that it does produce a scale invariant spectrum of perturbations, however the amplitude of the perturbations turns out to be proportional to the bounce energy density \cite{WE1}. This is problematic because the bounce density in LQC is of the order of Planck density, resulting in the amplitude becoming too large as compared to observations.  The feasibility of the matter-Ekpyrotic scenario in LQC to produce scale-invariant perturbations was first explored in \cite{CaiWE2014}, where it was shown that the presence of the Ekpyrotic field can solve the problem of having too large an amplitude as obtained in \cite{WE1}. But, the analysis of \cite{CaiWE2014} utilized the deformed algebra approach for the perturbations in the Ekpyrotic phase, and hence faced difficulties in analyzing the ultraviolet modes in the vicinity of the bounce. Thus one needed to make  several assumptions and approximations in order to arrive at qualitative results on the power spectrum and stopped short of providing a detailed numerical analysis due to the above mentioned difficulties with ultraviolet modes. Such an analysis does not permit the exploration of the detailed effects of the bounce in LQC on the power spectrum. We note that an analysis of the matter-Ekpyrotic scenario has been carried out also in \cite{haro1}, but only for the restricted case where the Ekpyrotic field is modelled by a scalar field having a constant equation of state. Unfortunately, the latter setting of constant equation of state does not capture some key elements of the original Ekpyrotic scenario where the equation of state varies because of the dynamics of the scalar field.

In the present manuscript, we provide a detailed numerical study of cosmological perturbations in the matter-Ekpyrotic scenario in LQC with dust and a general Ekpyrotic scalar field as the matter-energy content without making any assumptions on the equation of state or any approximations considered above. In contrast with \cite{CaiWE2014} which was based on deformed algebra approach, we use the dressed metric approach \cite{aan2013} for numerically analyzing the propagation of the comoving curvature perturbations through the phase of Ekpyrotic field domination. This allows us to overcome the difficulties encountered in analyzing ultraviolet modes in previous work \cite{CaiWE2014} by avoiding the Jeans instability arising from the imaginary sound speed during the superinflationary phase which occurs in deformed algebra approach. With a  full numerical control to analyze the power spectrum for any range of modes and to understand the effects of the quantum bounce, we consider perturbations that initiate in the matter-dominated phase of the contracting branch and evolve them numerically through the bouncing regime and  the expanding phase. Further, we  also analyze the effects of the duration of the Ekpyrotic phase, i.e. the regime with equation of state $w > 1$, on the amplitude, as well as the scale-invariant regime of the  power spectrum. The analysis is carried out by considering different initial conditions which change the period of dust versus Ekpyrotic field domination and help us understand the robustness of results, as well as to understand the effects of changing the parameters of the Ekpyrotic scalar field potential. From the numerical simulations of the power spectrum, we also consider the spectral index  and then compare it with the most recent observational data. Our analysis shows a matter-dominated phase sourced only by dust can not ensure a spectral index consistent with the observations.

The plan of the manuscript is as follows. In Sec. \ref{sec:background}, we study the general features of the background dynamics of the matter-Ekpyrotic scenario in LQC for different initial conditions set at the bounce and also comment on the effects of the changes in the parameters of the Ekpyrotic potential on the  phase of Ekpyrotic field domination, especially on the regime  with equation of state greater than unity in the contracting branch. In Sec. \ref{perturbations}, we consider quantum vacuum perturbations in the far past in the matter-dominated era of the contracting branch, and evolve them numerically to the expanding branch in the dressed metric approach of LQC. By analyzing the spectrum of perturbations taken at different times in both contracting and expanding phases, we show the properties of the power spectrum at different evolutionary stages and the effect of the bounce on the power spectrum.  Moreover,  based on the numerical results of the power spectrum, we also analyze the spectral index predicted by the model and  compare it with the observational data. Finally we end with concluding remarks in Sec. \ref{conclusion}.

In this manuscript, we use the Planck units with $\hbar=c=G=1$. In the formulas, we  keep the Newton's constant $G$ explicit, while in the numerical solutions $G$ is also set to unity. 

\section{Background dynamics of the matter-Ekpyrotic bounce scenario in LQC}
\renewcommand{\theequation}{2.\arabic{equation}}\setcounter{equation}{0}
\lb{sec:background}
For the matter-Ekpyrotic scenario  to be viable one needs a 
way to patch together a contracting and an expanding universe through a bounce in such a way that the scale invariant perturbations generated  in the contacting phase can propagate to the expanding branch without losing their scale invariance property. In LQC, a bounce  generically takes place  when the matter energy density reaches Planck scale due to the underlying quantum geometry effects \cite{APS,ASstatusreport}. Thus it provides a natural setting for studying the bouncing universe scenario as an alternative to the inflationary paradigm. In this section, we study the effective background dynamics of the spatially flat  homogeneous  and isotropic loop quantum cosmology with pressureless dust and a general Ekpyrotic scalar field as the matter content. We consider both dust and the Ekpyrotic field to be minimally coupled to gravity. Using the effective Hamilton's equations for the background dynamics, we numerically solve for the background solutions  with the initial conditions set at the bounce point.

\subsection{The effective dynamics of LQC in the matter-Ekpyotic bounce scenario}
LQC is based on the canonical quantization of the  symmetry-reduced cosmological models using techniques from loop quantum gravity (LQG) \cite{review, ASstatusreport}. In LQC, the classical Hamiltonian constraint is  reformulated in terms of the Ashtekar-Barbero connection and its conjugate triad, which in a spatially flat FLRW universe are symmetry  reduced to a canonical pair, namely $c$ and $p$. This reformulated Hamiltonian description of classical spacetimes is then quantized in the so-called $\bar \mu $ scheme, yielding a nonsingular quantum difference equation with equal steps in volume \cite{APS}.  In many practical applications of LQC to investigate the phenomenological implications of  quantum gravity effects on  the isotropic and anisotropic spacetimes, as well as on the linear perturbations around the background spacetimes,  it is always more convenient to make use of the effective dynamics of LQC \cite{as2017},  which has been shown  to faithfully represent the discrete quantum evolution of certain sharply peaked states in LQC for both isotropic and anisotropic spacetimes \cite{DGS2014a,DGS2014b,DJMS2017,ps18}. The effective dynamics is prescribed by an effective Hamiltonian constraint in a  phase space spanned by both gravitational and matter degrees of freedom. Due to the  homogeneity and  the isotropy of a spatially flat FLRW universe, the gravitational sector consist of  a canonical pair $(b,v)$, with   $v=|p|^{3/2}$ and $b=c|p|^{-1/2}$ and the fundamental Poisson bracket $\lbrace b,v \rbrace = 4\pi G \gamma$, where $\gamma$ is the  Barbero-Immirzi parameter. The latter's value is set by black hole thermodynamics and as in other works in LQC we will take this value to be $\gamma \approx 0.2375$ for numerical studies. The matter sector is composed of an Ekpyrotic field $\phi$ and its conjugate momentum $p_\phi$  with the fundamental Poisson bracket $\lbrace \phi,p_\phi \rbrace = 1$, as well as a dust field whose  energy density is denoted by $\rho_\mathrm{dust}$. In the above settings, the effective Hamiltonian describing the dynamics of loop quantized spatially flat FLRW model is given by \cite{APS}
\bq
\label{ham}
\mathcal{H}=-\frac{3v}{8\pi G \gamma^2 \lambda^2} \sin^2(\lambda b) + \mathcal{H_\mathrm{m}},
\eq
here $ \lambda = \sqrt{\Delta} $ with $ \Delta = 4 \sqrt{3} \pi \gamma \ell_{\mathrm{Pl}}^2 $ being the minimum area eigenvalue in LQG. $ \mathcal{H_\mathrm{m}} $  represents  the matter Hamiltonian,  which consists of a dust field and an Ekpyrotic field and hence takes the form
\bq
\lb{2a1}
\mathcal{H_\mathrm{m}}=\frac{{p_{\phi}}^2}{2v}+vU(\phi)+\mathcal E_\mathrm{dust},
\eq
where $\mathcal E_{dust}=\rho_{dust}v$ is the dust energy which remains a constant of motion. As proposed in \cite{CaiBB2012}, we take the Ekpyrotic potential $U(\phi)$ to be,
\bq
\lb{potential}
U(\phi)=\frac{-2u_o}{e^{-\sqrt{\frac{16\pi}{p}}\phi}+e^{\beta\sqrt{\frac{16\pi}{p}}\phi}},
\eq
with $u_o$, $p$ and $\beta$  all taking positive values. The Ekpyrotic potential (\ref{potential}) is negative definite and approaches zero as $\phi\rightarrow \pm \infty$. The potential has only one minimum at $\phi_\mathrm{min}=-\sqrt{\frac{p}{16\pi}}\ln \beta/(1+\beta)$, whose value is given by 
\bq
\lb{min}
U_\mathrm{min}=-\frac{2u_0}{1+\beta}\beta^{\frac{\beta}{1+\beta}}.
\eq
Moreover, the parameter $p$ also determines the width of the potential which increases with an increasing $p$. The potential is asymmetric about the potential minimum, and the degree of asymmetry depends on the choice of the parameter $\beta$.

From the effective Hamiltonian constraint, it is straightforward to derive the Hamilton's equations of motion, which read 
\bqn
\lb{2a2}
\dot b &=& - \frac{3 \sin^2(\lambda b)}{2 \gamma \lambda^2} -4\pi G\gamma P, \label{b_dot} \\
\lb{2a3}
\dot v &=& \frac{3 \sin(2\lambda b)}{2 \gamma \lambda} v, \label{v_dot} \\
\lb{2a4}
\dot \phi &=& \frac{p_{\phi}}{v}, \quad \quad 
\dot p_{\phi} =v\, U_{,\phi},
\eqn 
where $U_{,\phi}$ stands for the differentiation of the potential with respect to the Ekpyrotic field and $P$ is the isotropic pressure $ P=-\frac{\partial \mathcal{H_\mathrm{m}}}{\partial v} $.   From (\ref{2a3}), it can be shown that the modified Friedmann equation in LQC takes the form 
\bq
\lb{friedmann}
H^2 = \frac{8\pi G}{3} \rho \left(1-\frac{\rho}{\rho_{c}}\right),
\eq
where $ \rho_{c} = 3/(8\pi G \gamma^2 \lambda^2) \approx 0.41 \rho_{\mathrm{Pl}} $ is called the critical energy density in LQC. We note from the above equations that the Hubble rate is generically bounded. In a contracting universe, as the energy density increases and reaches a maximum $\rho_{c}$, the Hubble rate vanishes and reverses sign and a bounce occurs. This happens regardless of the type of matter content. The continuity equation for the matter fields, which also holds for each individual component, 
\bqn
\lb{continuity}
\dot\rho+3H(\rho+P)=0,
\eqn
remains unchanged with $\rho$ and $P$ given respectively by 
\bq
\lb{2a5}
\rho=\frac{{\dot \phi}^2}{2}+U(\phi)+\frac{\mathcal E_\mathrm{dust}}{v}, \quad P=\frac{\dot \phi^2}{2}-U(\phi). 
\eq
Note the dust field is pressureless and thus makes no contribution to $P$.

\subsection{Numerical analysis of the background dynamics}
In this subsection, using the effective Hamilton's equations, we numerically solve for solutions of the background dynamics in a spatially flat FLRW universe in LQC. We find that a non-singular bounce is obtained for all the solutions, and for various cases one finds multiple non-singular bounces in the Planck regime which separate two macroscopic regimes of a universe. 
Since our goal is to understand the way perturbations propagate from a macroscopic contracting branch to a macroscopic expanding branch we consider potential parameters such that there is only a single bounce in the studied time range. Since dust is pressureless and the Ekpyrotic potential is negative definite, it follows from \eqref{2a5}, that the total isotropic pressure is positive definite. Thus the dynamical variable $b$ is monotonic as $\dot b$ is negative definite according to \eqref{b_dot}. In view of \eqref{v_dot} for $\dot v$, this provides an easy way of calculating the number of bounces as $\left(b(t_i)-b(t_f)\right)\lambda /\pi$, where $t_i$ and $t_f$ are the initial and final times for the numerical evolution. For the time range $t\in(-10^6,10^6)$, we 
set the  parameters characterizing  the Ekpyrotic potential as  
\bq
\lb{parameters}
u_o=0.75, \quad p=0.10, \quad \beta=5.0,
\eq
this choice results in $\phi_\mathrm{min}=-0.012$ and $U_\mathrm{min}=-0.96$. 

The initial conditions for the numerical simulations are chosen at the bounce, where $t=t_B=0$. Hereafter, the index `B' refers to the values of the relevant quantities at the bounce point. In order to numerically solve the Hamilton's equations,  one needs to specify the initial values of $v_B$, $b_B$, $\phi_B$, $p_{\phi_B}$ and $\mathcal E_{dust}$.  Since the Hubble rate vanishes at the bounce, we choose $b_B=\pi / 2\lambda$ which corresponds to the branch that has the right classical limit as the matter density approaches zero. The bounce volume is set to $v_B=1$ (in Planck volume) in all cases without any loss of generality as the effective Hamilton's equations are invariant with respect to the rescaling of the volume and the momentum of the Ekpyrotic field. In this way, the parameter space at the bounce is made of two free  parameters, the Ekpyrotic field and the dust energy. With regard to the former, it is chosen to be initially positioned  near the minimum of the potential well at the bounce, namely $\phi_B=0$.  For the latter, we consider several different choices of  $\mathcal E_{dust}$ to investigate its impact on the background dynamics. Once $v_B$, $b_B$, $\phi_B$ and $\mathcal E_{dust}$ are specified, the value of $p_{\phi_B}$ at the bounce can be obtained from the vanishing of the Hamiltonian constraint (\ref{ham}) while the sign of  $p_{\phi_B}$ is specified so that the Ekpyrotic field is initially moving away from the bottom of the potential well.  

\begin{figure}
\includegraphics[width=8cm]{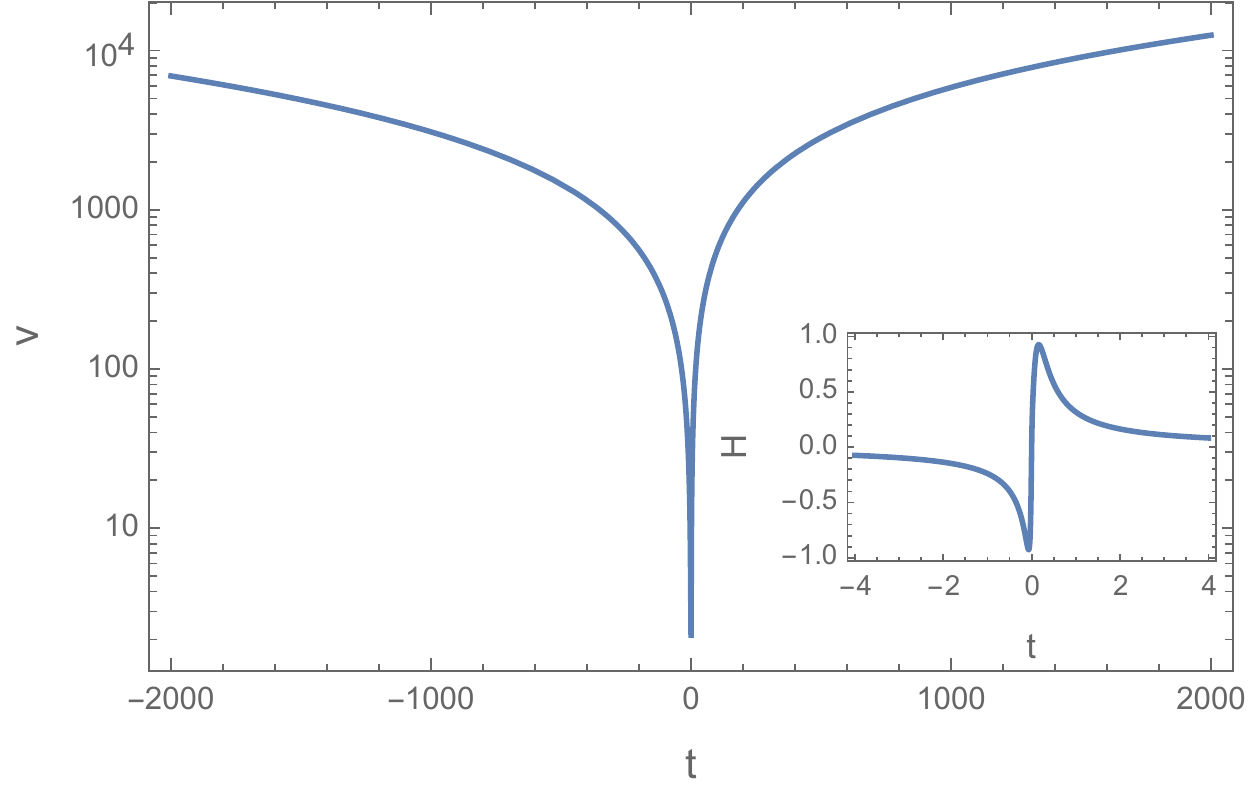}
\includegraphics[width=8cm]{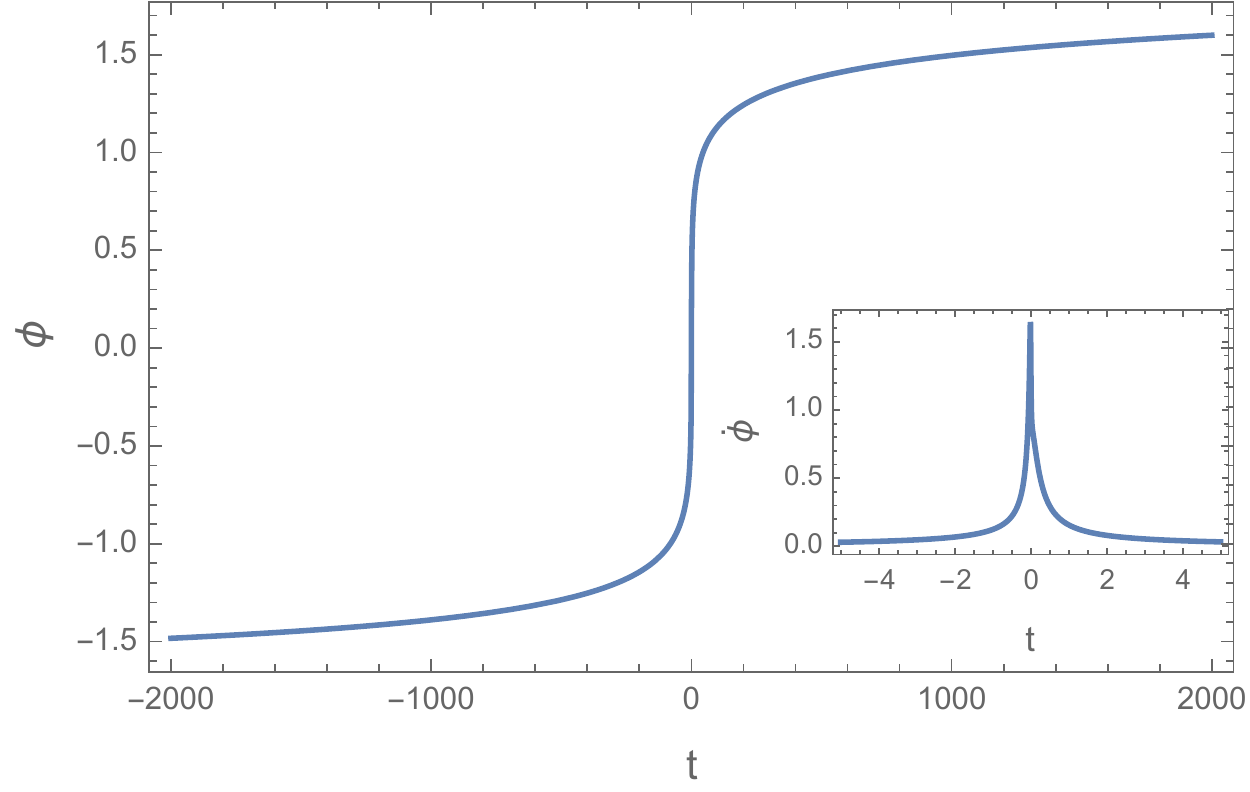}
\caption{The numerical evolution of the volume and the Ekpyrotic field are plotted for the initial conditions (\ref{initial1}). The inset plot in the left (right)  panel shows the behavior of the Hubble rate (the velocity of the Ekpyrotic field) near the bounce.}
\label{f1}
\end{figure}

\begin{figure}
\includegraphics[width=8cm]{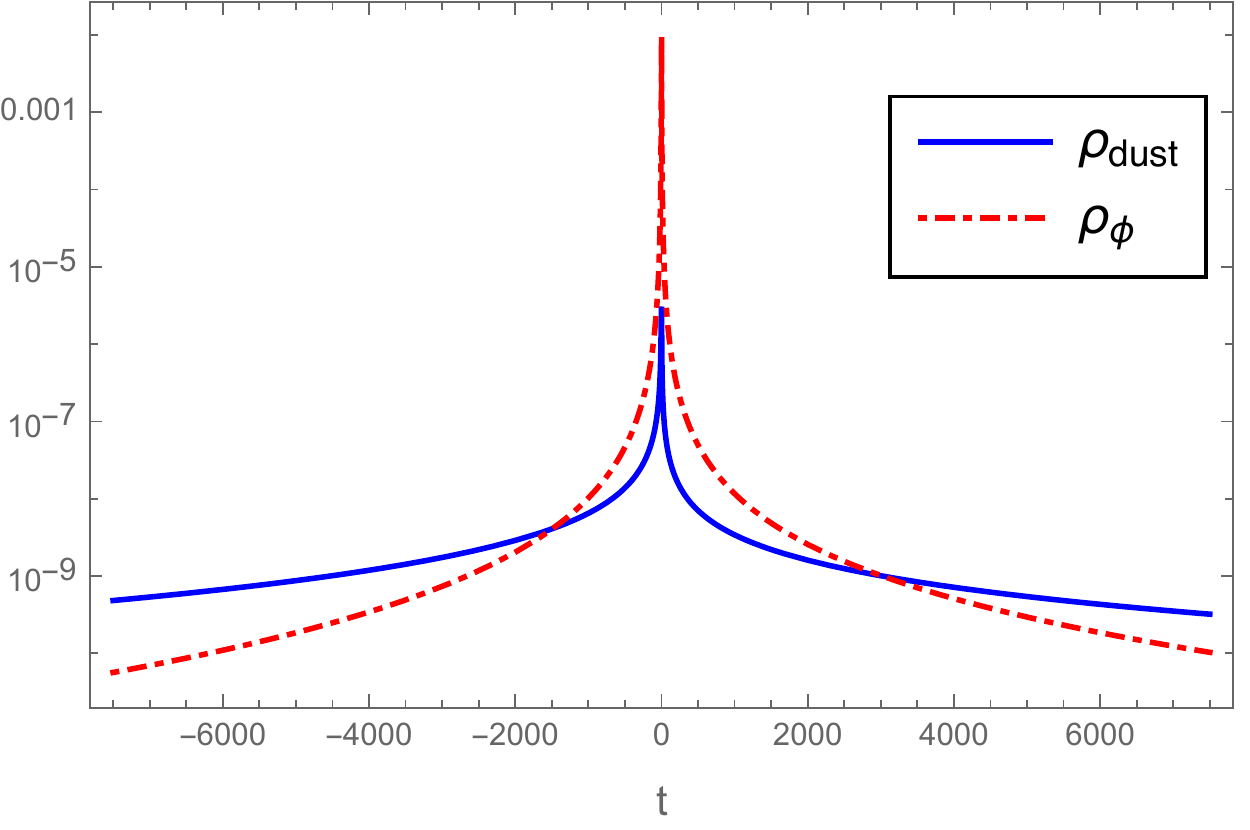}
\includegraphics[width=8cm]{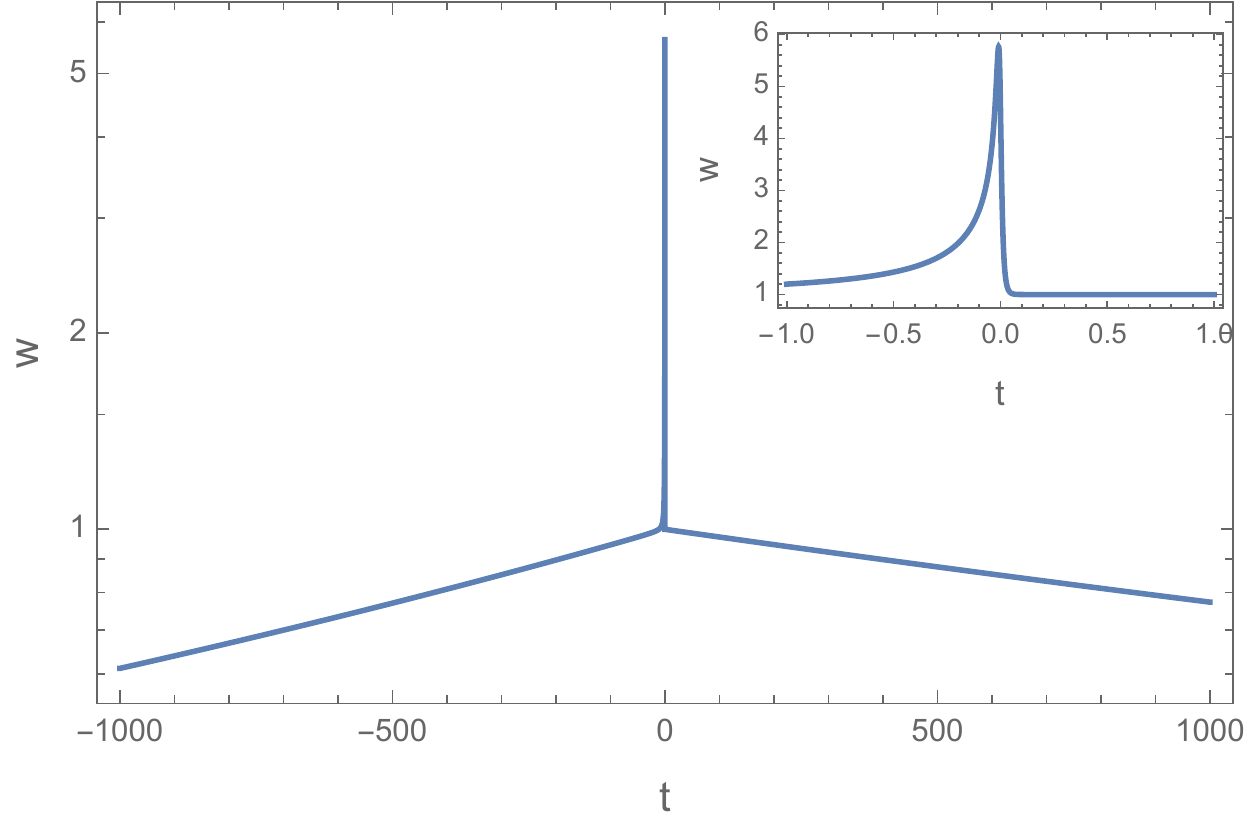}
\caption{In this figure, the evolution of each matter component and the equation of state are plotted for the initial conditions (\ref{initial1}). The energy density of the Ekpyrotic field overtakes the dust energy density at $t_1\approx-1.49\times10^3$ in the contracting phase. In the expanding phase, the dust field becomes dominant again after the transition point $t_2\approx3.02\times10^3$. The right panel shows the behavior of the equation of state when the Ekpyrotic field is dominant and the inset plot depicts the details of the change in the equation of state near the bounce point.   }
\label{f2}
\end{figure}

The first example  is presented in Figs. \ref{f1}-\ref{f2}  which correspond to the initial conditions
\bq
\lb{initial1}
\phi_B=0, \quad \quad \mathcal E_{dust}=2.00 \times 10^{-5}, \quad \quad p_{\phi_B}=1.50.
\eq  
As a result, at the bounce, the potential energy is $U_B=-0.75$ and the kinetic energy is $K_B=1.16$ (in Planck units). As compared with the Ekpyrotic field, the energy density of the dust field is very small at the bounce and the equation of state at the bounce turns out to be $w_B \approx 5.76$.  In Fig. \ref{f1}, the numerical evolution of the volume and the Hubble rate  shows a non-singular bounce at $t=0$. As the universe evolves smoothly from contraction to expansion, the Ekpyrotic field monotonically moves from the left wing of the potential to the right wing.  Fig. \ref{f2}  shows that the universe undergoes  two distinct phases during contraction, namely the matter-dominated phase and the  scalar field dominated phase, which are separated by the transition point at $t_1\approx-1.49\times10^3$. Before $t=t_1$, the energy density is dominated by the dust field. Afterwards, the Ekpyrotic field starts to dictate dynamics. We thus identify the scalar field dominated phase as the period from $t=t_1$ to the bounce point. A similar pattern can be observed  in the expanding phase where the transition point is  located at $t_2 \approx 3.02\times10^3$. Due to the asymmetry of the potential ($\beta\neq 1$), the  evolution of the universe is also asymmetric with respect to the bounce.  Note that the equation of state is not always greater than unity in the scalar field dominated phase. It is only in a small regime near the bounce that the equation of state becomes greater than unity as shown in the inset plot of Fig. \ref{f2}. This is mainly because  the regime with $w>1$, which as shown in Fig. \ref{f2} only lasts for about $1.1$ e-foldings, corresponds to the moment when the Ekpyrotic field is traversing the bottom of the potential. Since the Ekpyrotic potential is rather steep and narrow near its bottom, it takes a very short time for the Ekpyrotic field to move across the bottom of the potential, resulting in a short period with $w>1$. Note in  the Ekpyrotic scenarios,  the $w>1$ regime  corresponds to the Ekpyrotic phase, which in the current case can be prolonged a little bit  by increasing the width of the potential, namely increasing the value of $p$. For example, for $p=0.14$, the regime with $w>1$ in the contracting phase can last for $2.0$ e-foldings. 
In the expanding phase, as the Ekpyrotic field moves away from the bottom of the potential, the equation of state quickly drops below unity and decreases monotonically. The Ekpyrotic field remains the main component of the matter until the transition point $t_2$. Afterwards, the dust field starts to dominate again. 

\begin{figure}
\includegraphics[width=8cm]{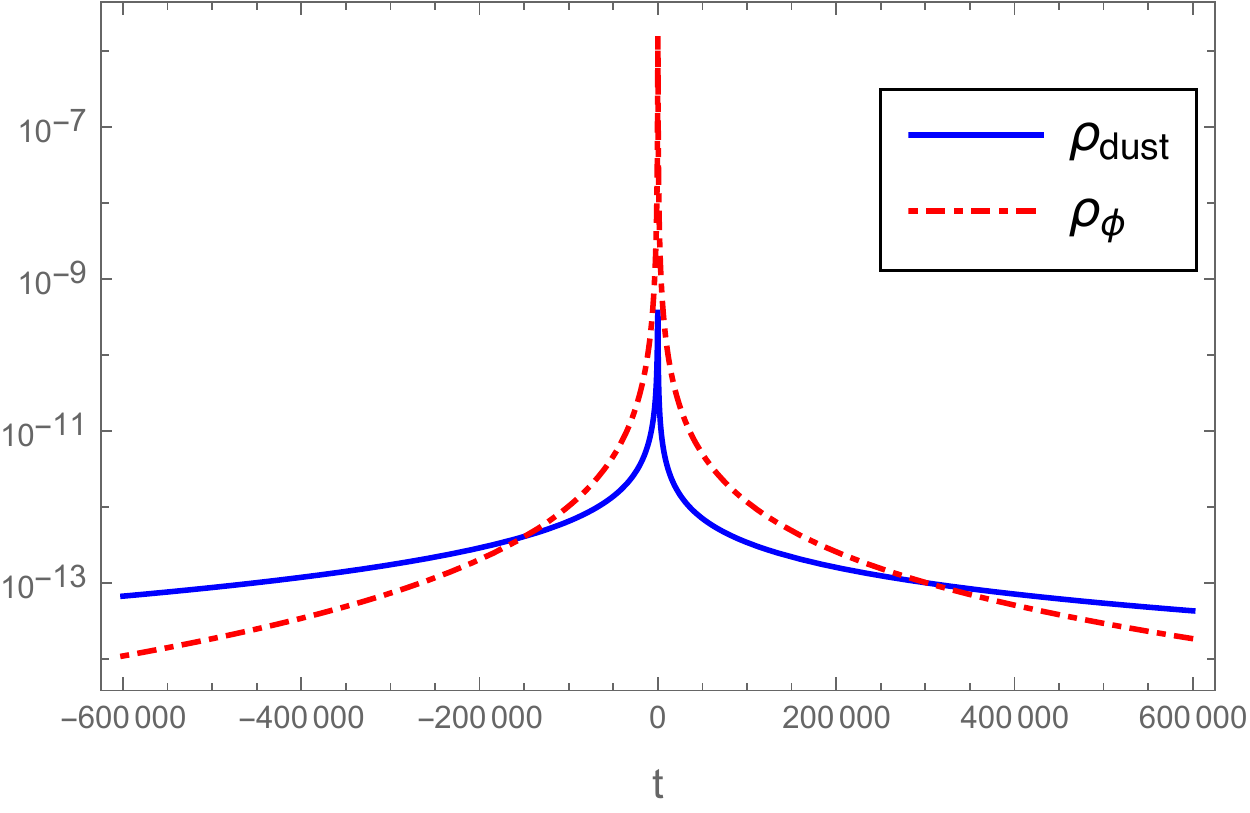}
\includegraphics[width=8cm]{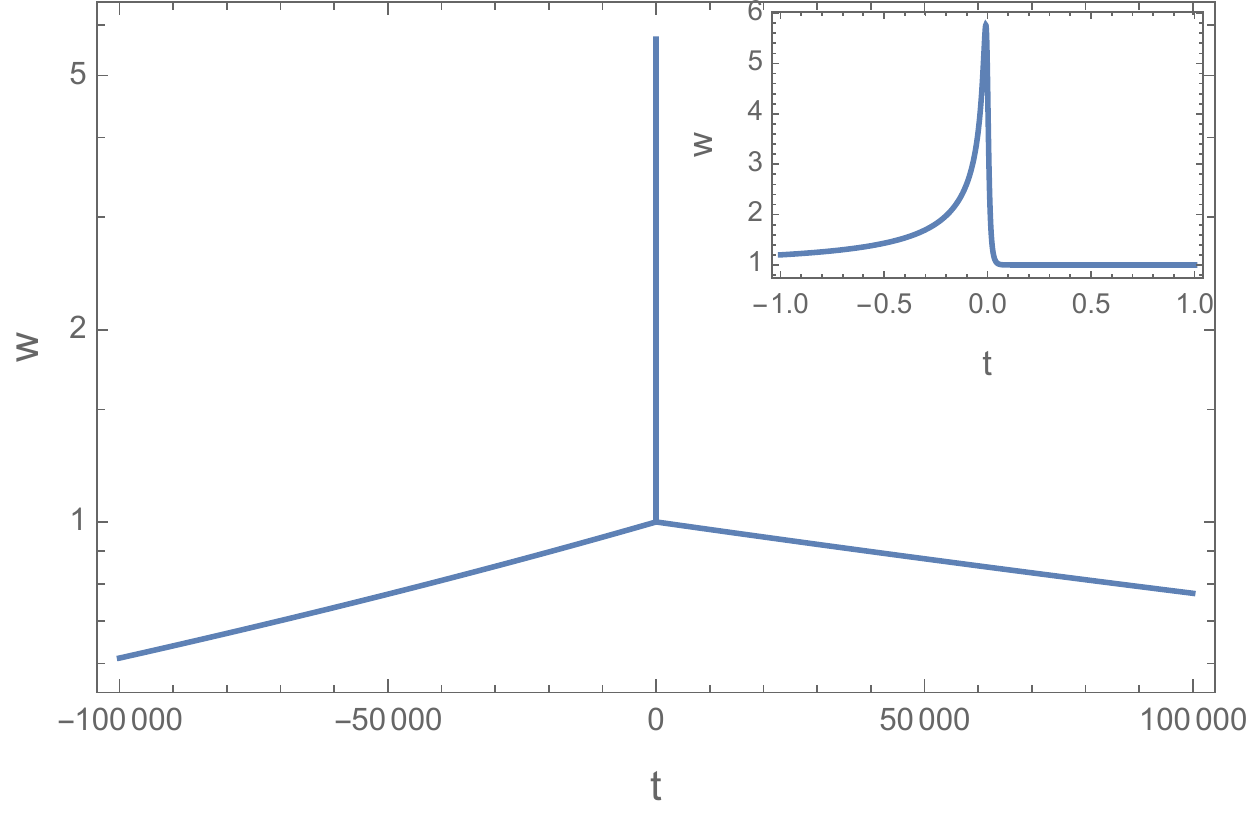}
\caption{The evolution of the energy density components and the equation of state are shown for the initial conditions (\ref{initial2}). The energy density of the Ekpyrotic field starts to become dominant after the transition point at $t_1\approx-1.49\times10^5$. In the expanding phase, there exists a second transition point at $t_2\approx3.02\times10^5$ where the universe enters into a matter-dominated  phase again. In the right panel, the equation of state approaches unity and goes above it for a very brief period just before the bounce, and then quickly declines below unity after the bounce. The inset figure in right plot shows the behavior of equation of state  near the bounce.  }
\label{fig}
\end{figure}

The second example given in Fig. \ref{fig} corresponds to the initial conditions
\bq
\lb{initial2}
\phi_B=0, \quad \quad \mathcal E_{dust}=2.00 \times 10^{-7}, \quad \quad p_{\phi_B}=1.50.
\eq  
In this case, we choose $\mathcal E_{dust}=2 \times 10^{-7}$ while keeping all the other parameters and the  initial conditions the same and the bounce is still dominated by the Ekpyrotic field.  Since the volume of the universe and the Ekpyrotic field evolve in  qualitatively the same way as in the last case, we only show the plot for the energy density and the equation of state in Fig. \ref{fig} in this case. When the dust energy  decreases, the duration of the scalar field dominated phase increases correspondingly as Fig. \ref{fig} shows that  the transition time  now becomes $t_1 \approx -1.49\times10^5$ in the contracting phase and $t_2 \approx 3.02\times10^5$ in the expanding phase. In particular,  the Ekpyrotic phase with $w>1$ in the contracting phase is also increased, which in the current case lasts $2.6$ e-foldings. Similar to the last case, one can prolong this regime by increasing the value of $p$. We find for $p=0.14$, the number of e-foldings for the regime with $w>1$ becomes $2.9$. Further increase in the value of $p$  causes multiple bounces in the given  time range.

\begin{table}
\caption{In the table, we list the transition time $t_1$ between the matter-dominated phase and the scalar field dominated phase in the contracting branch and the duration of the regime with $w>1$ for several different choices of $u_0$ when all the other parameters of the potential and $\mathcal E_{dust}$ are chosen the same as in (\ref{parameters}) and (\ref{initial2}).  Time is measured in Planck seconds. }
\begin{center}
 \begin{tabular}{||c | c| c||} 
 \hline
 $u_o$ & $t_1$ (contracting branch) & duration of $w>1$ regime (contracting branch) \\ [0.5ex] 
 \hline\hline
 1 & $-8.70 \times 10^{4}$ & 100 \\ 
 \hline
 0.75 & $-1.50 \times 10^{5}$ & 65 \\
 \hline
 0.03 & $-3.00 \times 10^{5}$ & 11 \\
 \hline
 0.008 & $-3.04 \times 10^{5}$ & 6.5 \\ [1ex] 
 \hline
\end{tabular}
\end{center}
\lb{table1}
\end{table}

Before the end of this section, we further analyze the impact of changing the parameter $u_o$ in the Ekpyrotic potential (\ref{potential}). It is found that while increasing $u_o$ decreases duration of the overall scalar field dominated phase on  one hand, it simultaneously increases the duration of the Ekpyrotic phase as a fraction of the scalar  field dominated phase. Recall that in our terminology, the latter phase implies the regime where Ekpyrotic field density dominates but the equation of state may not necessarily be $w > 1$ (which is a characteristic of a pure Ekpyrotic phase).  This is illustrated by the results shown in Table. \ref{table1}, where we have varied $u_o$ while keeping all the other parameters of the potential and $\mathcal E_{dust}$ the same as in (\ref{parameters}) and  (\ref{initial2}). All these cases have only one bounce in the plotted range $t\in(-10^6,10^6)$.
This effect due to change in $u_o$ can be understood as follows. While increasing $u_o$ increases the steepness of the potential well, it also increases the depth of the potential well. Consider the case when we double $u_o$. This doubles the potential energy of the Ekpyrotic field at bounce, compared to what it was before. Since the potential is negative definite, it can be seen from the expression for energy density \eqref{2a5} that doubling $u_o$ while keeping $\mathcal E_{dust}$ fixed at fixed bounce volume has the effect of increasing the kinetic energy at the bounce (in order to ensure that the total energy density at bounce still adds up to $\rho_c$ -- the critical density at bounce in LQC). Note that the Ekpyrotic energy density at bounce is still the same as before since $\rho_c$ and $\rho_{dust}$ at bounce are unchanged. From the equation \eqref{2a5}, we see that pressure also increases compared to before. Thus, while the Ekpyrotic energy density at bounce is the same as before, the kinetic energy and pressure at bounce have substantially increased (substantially increasing the equation of state at the bounce as well). This means that the field climbs up the well faster, but due to a deeper well, the equation of state $w$ is larger than $1$ for most of this climb. Contrast this with the effect of increasing $p$ while keeping everything else fixed - it only decreases the steepness of the well while the depth of the well is unchanged - thus the kinetic energy, potential energy, pressure and energy density of the Ekpyrotic field at the bounce are all unaffected, except that the well is now less steep. Another effect of increasing $u_o$ is that it lowers the threshold for $p$ to get more bounces in a given time range.

In summary, from the numerical analysis of the background dynamics, we observe a brief period right before the bounce in the contracting phase in which the equation of state becomes greater than unity (namely, the Ekpyrotic phase). Decreasing the dust energy density, or increasing the width of the Ekpyrotic potential (by increasing $p$), or increasing $u_o$ can help increase the duration of such a period. As discussed above, the effects of these changes also differ qualitatively from one another in other details. However, the number of the bounces in a certain given time range is also sensitive to the shape of the potential. We find an appropriate region in the parameter space of the Ekpyrotic potential as given in (\ref{parameters}), where the duration of the Ekpyrotic phase (the regime with $w>1$) and the scalar field dominated phase can be varied while maintaining a single bounce in  the given range $t\in(-10^6,10^6)$.

\section{Scalar power spectrum in the matter-Ekpyrotic bounce scenario with the dressed metric approach}
\renewcommand{\theequation}{3.\arabic{equation}}\setcounter{equation}{0}\lb{perturbations}

In this section, we discuss the scalar power spectrum from the matter-Ekpyrotic bounce scenario in the framework of  LQC. From the numerical analysis of the background dynamics in the last section, we know there are two distinct phases in the contracting branch before the universe reaches the bounce  point, namely, the matter-dominated phase  dominated by dust and the scalar field dominated phase dominated by the Ekpyrotic field. In the former, the total energy density is far below the Planck density so that the background dynamics is essentially governed by the classical Friedmann equation, while in the latter, the modifications to Friedmann dynamics obtained from the effective spacetime regime become important since the scalar field dominated  phase  overlaps with the bouncing regime where the energy density becomes Planckian.  When it comes to the quantum perturbations around the background spacetime, a rigorous treatment of the linear perturbations should consider the perturbations of the dust field and the Ekpyrotic field and then construct the gauge-invariant perturbations which corresponds to the comoving curvature perturbation and the entropy perturbation. In the following, we only focus on the comoving curvature perturbation $\mathcal R_k$ which is  related to the Mukhanov-Sasaki  variable  $\nu_k$ via  $\nu_k=z\mathcal R_k$ with $z=a\dot \phi/H$. 

\subsection{The matter-dominated phase in the contracting phase}
We start with the matter-dominated phase where the dust field is dominant over the Ekpyrotic field. In this phase, one can find the approximate analytical solutions of the background dynamics by ignoring the contributions from the Ekpyrotic field. Note that the total energy density is far below the Planck density in the matter-dominated phase,  the modified Friedmann equation is well approximated by its classical counterpart
\bq
\lb{3a1}
H^2=\frac{8\pi G}{3}\rho.
\eq
Substituting $\rho=\mathcal E_\mathrm{dust}/v$ into the above equation, one can in a straightforward way  solve for the scale factor, which turns out to be 
\bq
\lb{3a2}
a=\left(a^{3/2}_i-\sqrt{6\pi G \mathcal E_\mathrm{dust}}(t-t_i)\right)^{2/3},
\eq
here $t_i$ is the initial time when the dust field is dominant and $a_i$ is the value of the scale factor at $t=t_i$. As the quantum gravity effects are negligible in this phase,   the perturbation equation  in terms of the  Mukhanov-Sasaki  variable also takes its classical form 
\bq
\lb{3a3}
\nu''_k+\left(k^2-\frac{z''}{z}\right)\nu_k=0,
\eq
where the prime represents the differentiation with respect to the conformal time. Using the relation $d\eta=dt/a$,  one can express the scale factor in terms of the conformal time and obtain
\bq
\lb{3a4}
a=\frac{2\pi G \mathcal E_\mathrm{dust}}{3}\left(\eta-\eta_i-\frac{3}{\sqrt{6\pi G \mathcal E_\mathrm{dust}}}a^{1/2}_i\right),
\eq
with $\eta_i$ denoting the conformal time corresponding to $t=t_i$. Noticing $z''/z=a''/a$ in the dust dominated phase, one is able to find the explicit form of the perturbation equation, which reads 
\bq
\lb{perturbation1}
\nu''_k+\left(k^2-\frac{2}{(\eta-\eta_0)^2}\right)\nu_k=0,
\eq 
where 
\bq 
\eta_0=\eta_i+\frac{3}{\sqrt{6\pi G \mathcal E_\mathrm{dust}}}a^{1/2}_i.
\eq
Note that the above perturbation equation in the matter-dominated phase takes the same form as the one in the inflationary de Sitter  phase which is  the basis of the duality  between a matter-dominated contracting phase and an inflationary spacetime  as discussed in \cite{Wands1999}.
The general solution of Eq. (\ref{perturbation1}) reads 
\bq
\lb{3a5}
\nu_k=\frac{\alpha_k e^{-ik(\eta-\eta_0)}}{\sqrt{2k}}\left(1-\frac{i}{k(\eta-\eta_0)}\right)+\frac{\beta_k e^{ik(\eta-\eta_0)}}{\sqrt{2k}}\left(1+\frac{i}{k(\eta-\eta_0)}\right),
\eq 
where two integration constants $\alpha_k$ and $\beta_k$  can be uniquely  fixed via the Wronskian condition \cite{DB09}
\bq
\lb{wronskian}
\nu_k\left(\nu^*_k\right)'-(\nu^*_k )\nu'_k=i, 
\eq
and the boundary condition 
\bq
\lim_{\eta\rightarrow -\infty} \nu_k=\frac{e^{-ik\eta}}{\sqrt{2k}}.
\eq
In this way, we choose the positive frequency states  $\alpha_k=1$ and $\beta_k=0$ for all the comoving wavenumbers  whose asymptotic states at $\eta \rightarrow -\infty$ are the Bunch-Davies (BD) vacuum. It should be noted that the solution  (\ref{3a5}) only holds with good approximation in the regime where the equation of state $w\approx 0$. As the universe evolves towards the bounce point in the contracting phase, the Ekpyrotic field would become dominant  and the total energy density increases towards the Planck scale. Therefore,  in the scalar field dominated phase, the perturbation equation  (\ref{perturbation1}) is neither  valid nor is the solution given in (\ref{3a5}). One is thus forced  to consider the perturbation theory in a quantum  spacetime where quantum gravity effects become important. Among all the approaches to the perturbation theory in LQC, we appeal to the dressed metric approach \cite{aan2013, lsw2020} for studying  the  propagation of the quantum perturbations across the bounce in order to avoid the Jeans instability encountered in the deformed algebra approach \cite{bgsl2015}. In the effective description of the quantum dynamics in the dressed metric approach, the evolution equation of the Mukhanov-Sasaki   variable  has the same form as their classical counterpart with the evolution of the  background variables  obeying the modified Friedmann equation instead of the classical Friedmann equation. As a result, this approach  is well suited for the purpose of the investigations on  how the  comoving curvature perturbations propagate  through the scalar field dominated phase near the bounce.

\subsection{The scalar field dominated phase near the bounce}
In the dressed metric approach, the quantum perturbations are described as propagating on a quantum spacetime which can be well approximated  by a differential manifold with a  dressed metric for the sharply-peaked semi-classical states.  The  evolution equation of the  Mukhanov-Sasaki   variable  takes the form \cite{lsw2020}
\bq
\lb{3b1}
\nu''_k+\left(k^2+\Omega^2-\frac{a''}{a}\right)\nu_k=0,
\eq
where $\Omega^2$   only depends on the background quantities and is  explicitly given by 
\bq
\lb{3b2}
\Omega^2=3\kappa  \frac{p^2_\phi}{a^4}-18\frac{p^4_\phi}{a^6\pi^2_a}-12a\frac{p_\phi}{\pi_a}U_{,\phi}+a^2U_{,\phi\phi},
\eq
here $\kappa=8\pi G$ and  $\pi_a$ is the conjugate momentum of the scale factor. In the effective description of the dressed metric approach, the relevant background quantities in the above equation are determined from the solutions of the modified Friedmann equation (\ref{friedmann}). Meanwhile, based on the exact form of the zeroth-order constraint, different forms of $\Omega^2$ can be reached. If one uses the classical Friedmann constraint and  expresses the $\pi_a$ in terms of the remaining variables, we can end up with 
\bq
\lb{typea}
 \Omega^2=a^2 \left( U_{,\phi \phi}+2 \cos\left(\lambda b\right) f U_{,\phi}+f^2 U\right),
\eq 
with  $f=\sqrt{24\pi G/\rho}\dot \phi$.  Moreover, $\cos\left(\lambda b\right) $ in the second term of the parenthesis comes from  the requirement to smooth $ \Omega^2$ across the bounce point. On the other hand,  since the background dynamics obeys the modified Friedmann equation in the effective approach, it is natural to use the effective Hamiltonian constraint to make replacement of $\pi_a$ in the expression of $ \Omega^2$ given in (\ref{3b2}). This ansatz is equivalent to making the replacement \cite{lsw2020}
\bqn
\lb{3b3}
\frac{1}{\pi^2_a}&\rightarrow& \frac{16\pi^2 G^2 \gamma^2\lambda^2}{9a^4\sin^2\left(\lambda b\right)}, \\
\lb{3b4}
\frac{1}{\pi_a}&\rightarrow& \frac{-4\pi G \gamma \lambda \cos\left(\lambda b\right)}{3a^2\sin\left(\lambda b\right)},
\eqn
in (\ref{3b2}) and the resulting $\Omega^2$ is denoted by $\Omega^2_\mathrm{eff}$.
The evolution equation (\ref{3b1}) reduces to (\ref{perturbation1}) in the matter-dominated phase for a massless scalar field when the matter density and the curvature is far below the Planck scale. 
In terms of the mode function $\nu_k$, the power spectrum of the comoving curvature perturbation is given by 
\bq
\lb{3b4}
\mathcal P_{\mathcal R}=\frac{k^3}{2\pi^2}\frac{|\nu_k|^2}{z^2},
\eq
with $z=a\dot \phi/H$.  In the matter-dominated phase, the superhorizon modes behave as $\nu_k \propto 1/k^{3/2}$ which implies these modes are  already scale invariant once they exit the horizon. As a result, the scalar field dominated phase, including the Ekpyrotic phase,  plays an important role in determining if the scale invariance of the  power spectrum would be preserved throughout the bouncing regime where quantum gravity effects come into play.  We analyze this important aspect in the next section with detailed numerical results of the power spectrum.

\subsection{Numerical results of the scalar power spectrum}

\begin{figure}
\includegraphics[width=8.1cm]{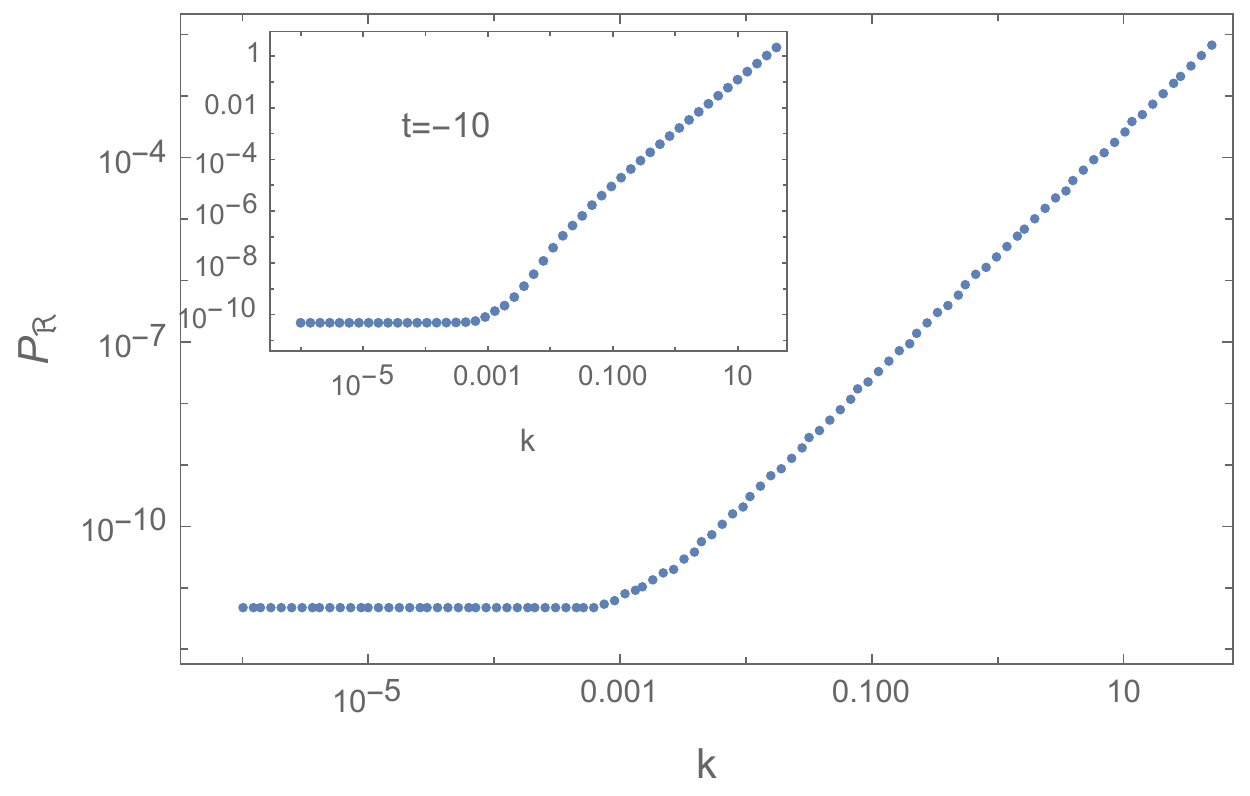}
\includegraphics[width=8cm]{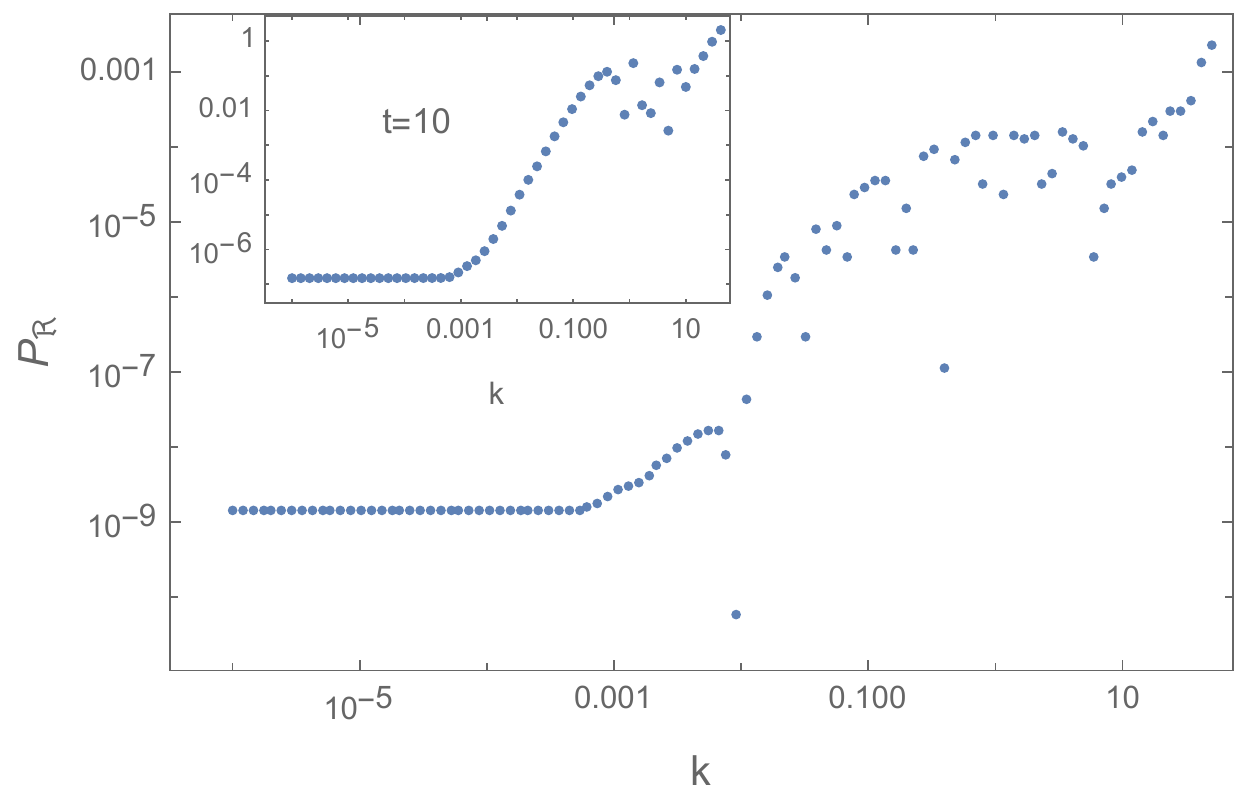}
\caption{With the initial conditions chosen as in (\ref{initial1}),  the power spectra at different times are depicted: in the left panel, the power spectra are evaluated at the transition time in the contracting phase ($t\approx-1.49\times10^3$) and $t=-10$ (in the inset plot) while in the right panel, the power spectra are evaluated at the transition time in the expanding phase ($t\approx3.02\times10^3$) and $t=10$ (in the inset plot). }
\label{f3a}
\end{figure}

\begin{figure}
\includegraphics[width=8.3cm]{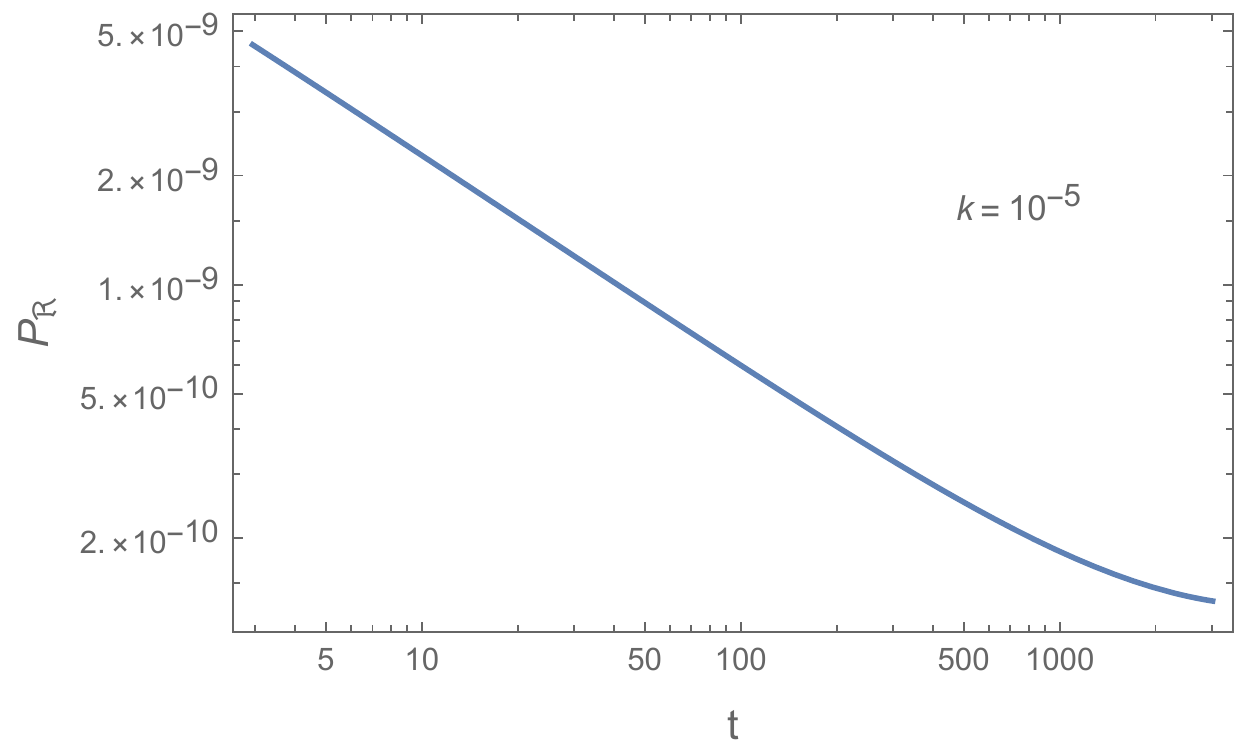}
\includegraphics[width=7.9cm]{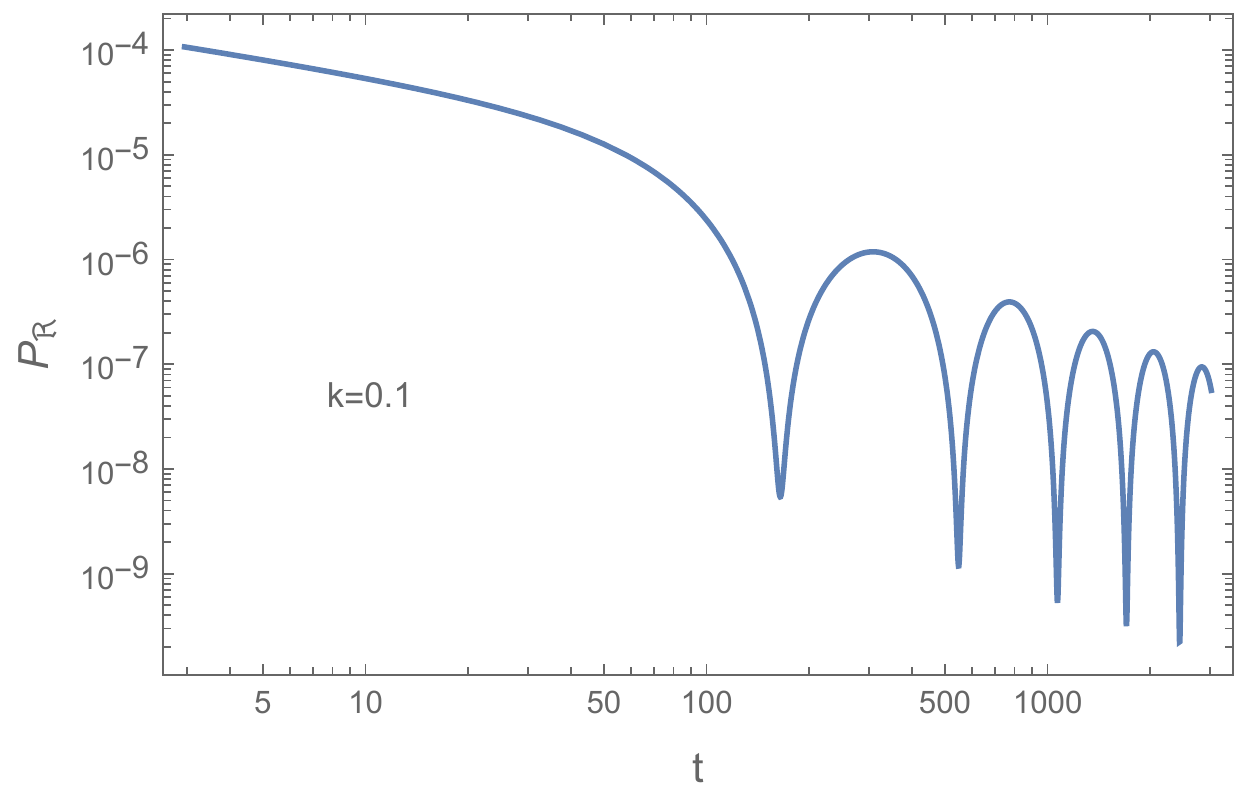}
\caption{With the same initial conditions as in Fig. \ref{f3a}, the power spectra for two different modes  are shown  in the expanding phase until the transition time when the energy densities of the dust field and the Ekpyrotic field become equal. The mode in the left panel is located in the scale invariant regime while the qualitative behavior of the mode in the right panel is affected by the bouncing regime. }
\label{f4a}
\end{figure}

In this subsection, we present the numerical results of the  scalar power spectrum for two separate cases of background dynamics studied in Sec. \ref{sec:background}. These two cases differ by the dust energy density at the bounce and the duration of the Ekpyrotic phase. In the following, we focus on the impact of the Ekpyrotic field and the bouncing phase on the evolution of the comoving curvature perturbations  until the transition point in the expanding phase. 

The first case is presented in Figs. \ref{f3a}-\ref{f4a} where the initial conditions for the background dynamics are chosen to be the same as in (\ref{initial1})  and the initial states for the scalar perturbations are set to (\ref{3a5}) with $\alpha_k=1$ and $\beta_k=0$. In Fig. \ref{f3a},   the  main plot in the left panel presents the scalar power spectrum evaluated at the transition point $t\approx -1.49\times10^3$ in the contracting phase,  which  shows that the scale invariant regime of the power spectrum is already generated at the end of the matter-dominated phase. In order to check if this regime is preserved in the contracting phase,  we also plot the power spectrum at a different time before the bounce, which is chosen to be  $t=-10$ (one can also choose any moment that is close to the bounce point). The resulting power spectrum is shown in the inset plot of the left panel.  It turns out that the magnitude of the power spectrum changes over time, which is in contrast with the constant power spectrum predicted by the analytical approximations for constant Ekpyrotic equation of state in the earlier work \cite{CaiWE2014}. As shown in Fig. \ref{f2} and Fig. \ref{fig} from Sec. \ref{sec:background}, the equation of state does not stay constant and varies during the entire evolution including during the scalar field dominated phase and the $w>1$ regime. Thus we find that the amplitude of the power spectrum is time-dependent. Moreover, the scale invariant regime lies in the same range of the comoving wavenumbers, namely $k\lesssim 10^{-3}$,  in the main plot and the inset plot. Similarly, we also present the scalar power spectrum at the transition point $t\approx3.02\times10^3$ in the expanding phase in the main plot of the right panel, which shows the same scale-invariant regime as in the contracting phase but with a different magnitude.  In addition, there is an oscillatory regime starting from $k\ge10^{-2}$ which is reminiscent of the oscillatory regime of the power spectrum produced in the inflationary scenario \cite{lsw2020}.   In order to show how this oscillatory regime appears at different times in the expanding phase, we plot the power spectrum at $t=10$ (again one can choose any other time after the bounce) in the inset plot of the right panel which shows an oscillatory regime for a different range of the comoving wavenumber. Apparently,  the power spectrum near $k=0.1$ already becomes oscillatory in the main plot while it is still monotonic in the inset plot.  This implies different modes become oscillatory at different times  in the expanding branch as they enter Hubble horizon at different times. Finally, we find the magnitude of the scale-invariant regime changes more rapidly near the bounce as can be seen by comparing the inset plots and the main plots in the figure. 

To explicitly show the evolution of different modes after the bounce, we study two distinct  modes,  namely  $k=10^{-5}$ and $k=0.1$ in Fig. \ref{f4a}. The first mode ($k=10^{-5}$) lies in the scale-invariant regime and decreases monotonically from the bounce point to the transition point. On the other hand, the second mode ($k=0.1$) becomes oscillatory after $t\approx100$ with a decreasing averaged magnitude. Based on the above analysis, we conclude that the scale invariance of the power spectrum which lies in the regime $k\lesssim 10^{-3}$ is preserved once it is generated at the end of the matter-dominated phase. Unlike the constant amplitude of the power spectrum in the scale-invariant regime generated during the inflationary phase, the magnitude of the power spectrum in the scale-invariant regime changes over time in the matter-Ekpyrotic bounce scenario. Again, we want to emphasize that our numerical results show that the amplitude of the power spectrum of the scale invariant regime is varying with time, which is in contrast with the analytical approximations in the earlier work where a constant equation of state was assumed for the Ekpyrotic phase \cite{CaiWE2014}. Further, while the bouncing regime does not change the character of the scale-invariant regime, it does have an impact on the power spectrum by producing an oscillatory regime  which does not overlap with the scale-invariant regime of power spectrum.

\begin{figure}
\includegraphics[width=8cm]{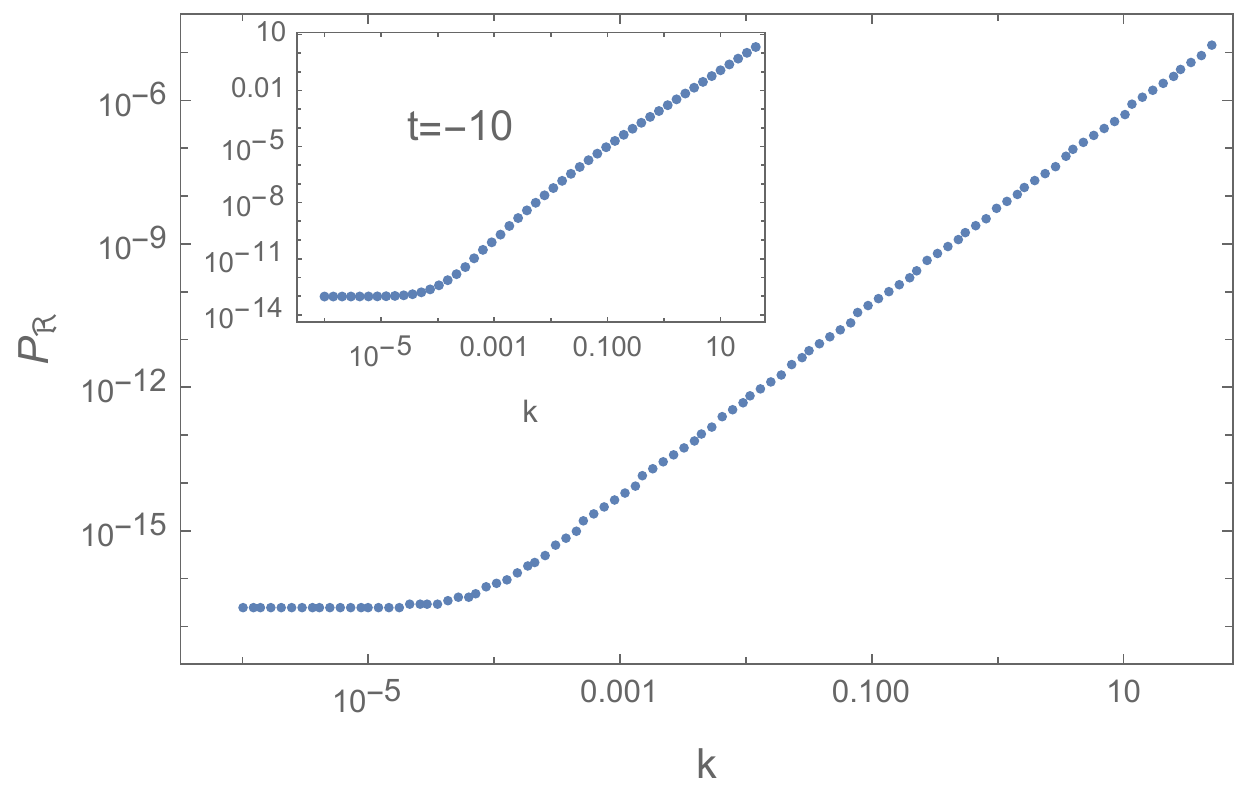}
\includegraphics[width=8cm]{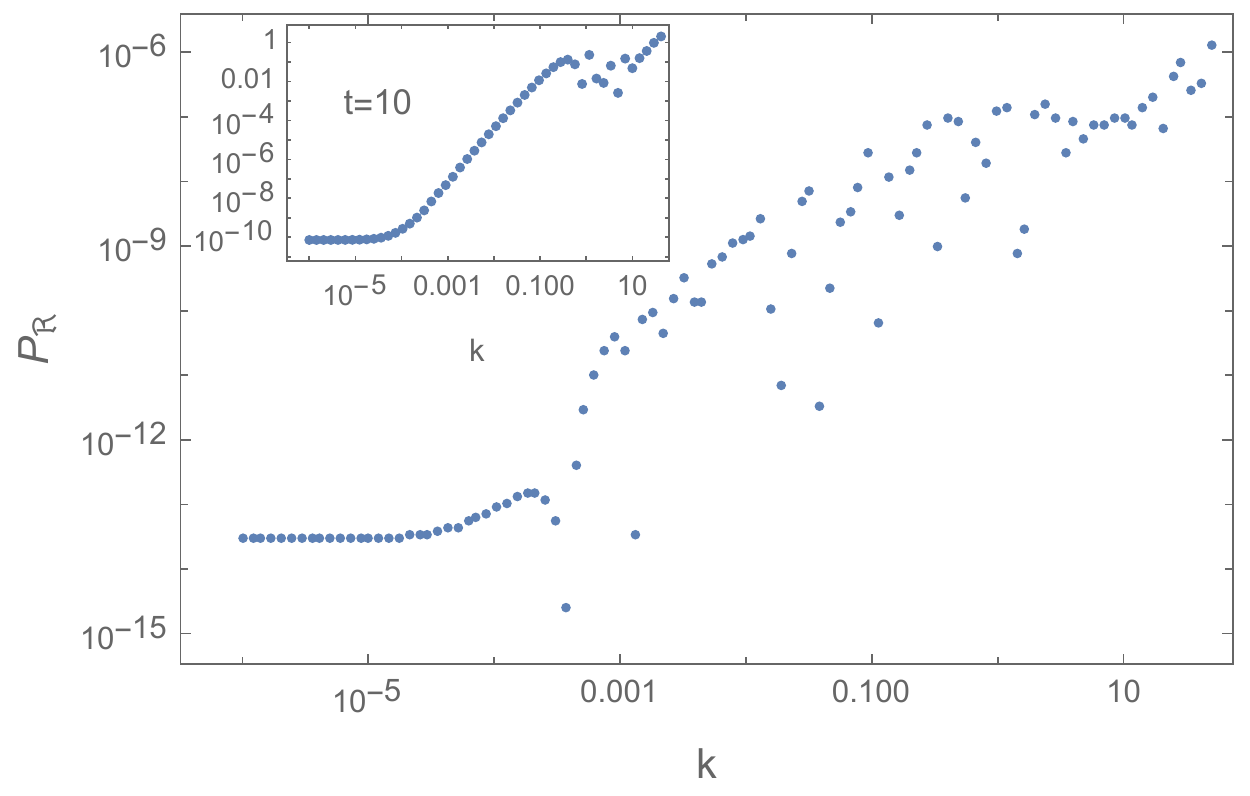}
\caption{As compared with Fig. \ref{f3a}, we change the dust energy to $\mathcal E_\mathrm{dust}=2.00\times10^{-7}$ and evaluate the power spectra at the transition time in the contracting phase ($t\approx-1.49\times10^5$) and the expanding phase ($t\approx 3.02\times10^5$), as well as the times near the bounce which are chosen to be $ t=-10$ (the inset plot in the left panel) and $ t=10 $ (the inset plot in the right panel). }
\label{f5a}
\end{figure}

\begin{figure}
\includegraphics[width=8.2cm]{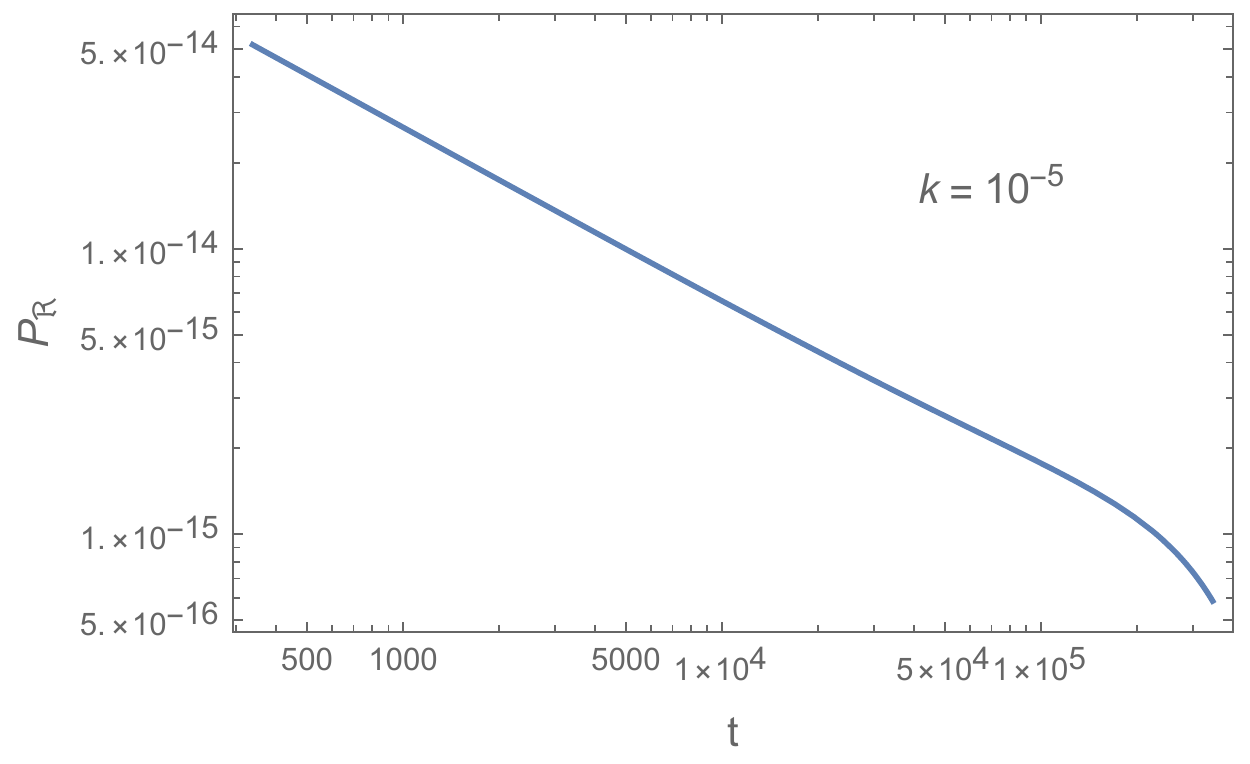}
\includegraphics[width=7.9cm]{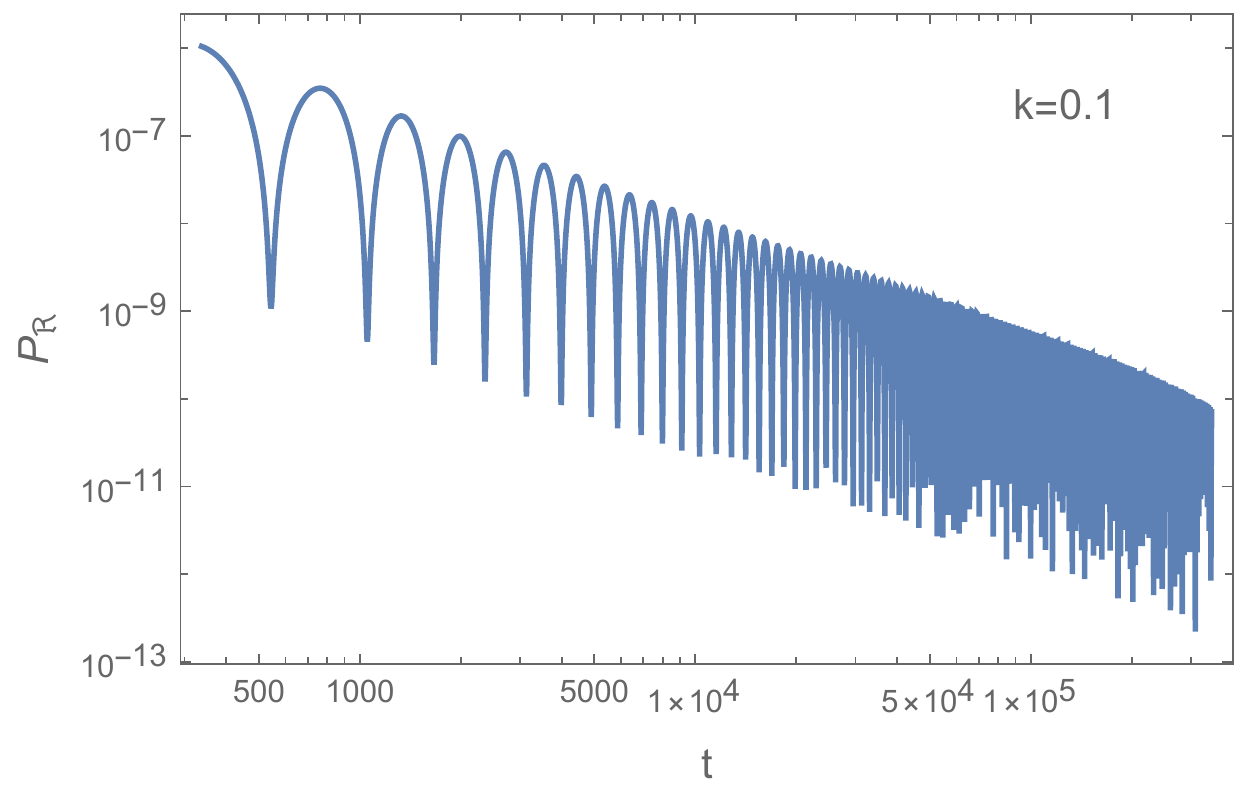}
\caption{With the same initial conditions as in Fig. \ref{f5a},  the evolution of the power spectra for two different modes are depicted in the expanding phase when the energy density of the  Ekpyrotic field is larger than the energy density of the dust field. These two modes exhibit different behavior during evolution as they are located in two separate regimes of the scalar power spectrum.  }
\label{f6a}
\end{figure}

The next example is presented in Figs. \ref{f5a}-\ref{f6a} where the initial conditions of the background dynamics  are set to (\ref{initial2}).  Fig. \ref{f5a} is plotted to compare  with Fig. \ref{f3a} while Fig. \ref{f6a}  is analogous to Fig. \ref{f4a}.  With a decrease in dust energy density, the duration of the scalar field dominated  phase, as well as the Ekpyrotic phase,  becomes longer.  As a result,   although the qualitative behavior of the power spectrum in each subfigure and the inset plot resembles  those  plotted in Figs. \ref{f3a}-\ref{f4a}, we find a smaller amplitude of the scale invariant regime of the power spectrum as compared with the last case. Moreover, the scale-invariant regime is now  located at $k\lesssim10^{-4}$ whereas  the oscillatory regime  shown in the main plot of the right panel of Fig. \ref{f5a} is extended to $k\gtrsim 10^{-3}$. This is because the scale invariance is exhibited by only those modes which exit the horizon during matter dominated contraction. As the duration of the scalar field dominated phase is increased, the modes around $k\approx10^{-3}$ now exit during the scalar field dominated phase and do not show scale invariance. This may be used to put yet another constraint on the duration of the Ekpyrotic phase using observations. Similar to the previous case, different modes in the oscillatory regime start to oscillate at different times in the expanding phase, which can be seen from a different behavior of the modes near $k=0.001$ in the main plot and the inset plot in the right panel of Fig. \ref{f5a}.  Meanwhile, the evolution of two representative modes $k=10^{-5}$ and $k=0.1$ depicted in Fig. \ref{f6a} have similar qualitative behavior as in the previous case with the former decreasing monotonically in the expanding phase and the latter becoming oscillatory after $t\approx100$.  

\begin{figure}
\includegraphics[width=8cm]{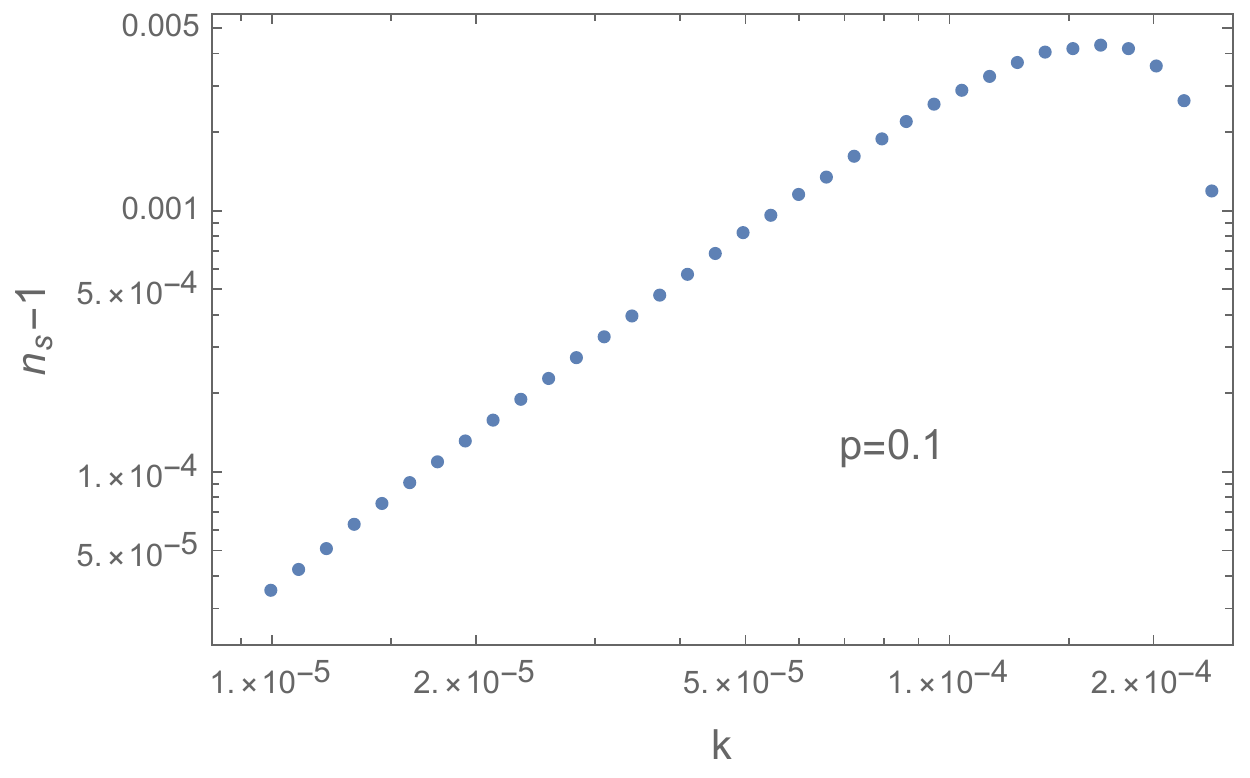}
\includegraphics[width=7.6cm]{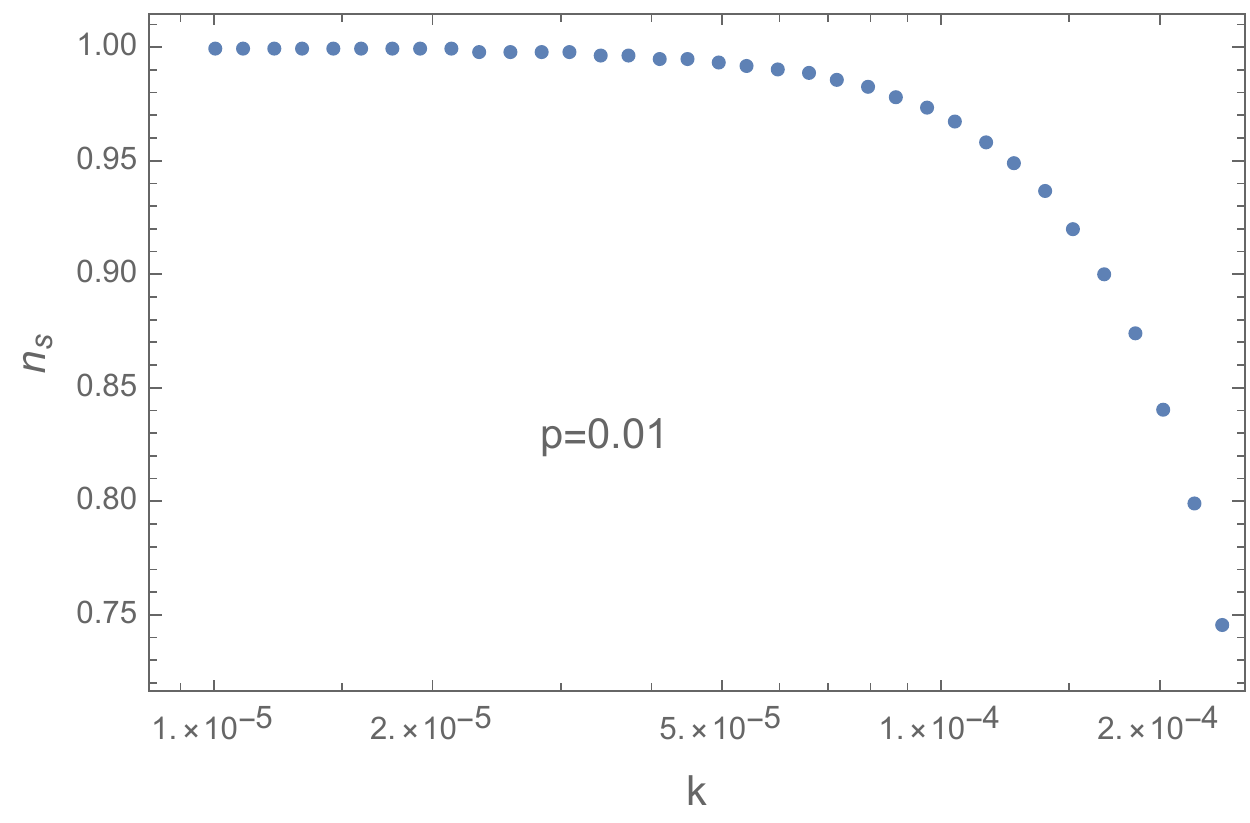}
\caption{For the power spectrum displayed in Fig. \ref{f3a}, we evaluate  the scalar index $n_s$ at the transition point $t\approx3.02\times10^3$ in the left panel and find $n_s$ is larger than unity in the scale invariant regime of the power spectrum. In the right panel, we only decrease the width of the potential while keeping other parameters unchanged. For the same initial values of the dust energy and the Ekpyrotic field,  we  find that $n_s$ now becomes less than unity in the same regime of the comoving  wavenumbers depicted in the left panel.}
\label{f7a}
\end{figure}

Finally, we discuss the spectral index $n_s$ in the matter-Ekpyrotic bounce scenario. In the current setting, the matter dominated phase is generated by an overwhelming dust field. We compute the spectral index in the expanding phase using our numerical results of the scalar power spectrum at the transition point in the expanding phase, and using 
\bq
n_s=1+\frac{\mathrm{d}\ln P_\mathcal R}{\mathrm{d} \ln k}.
\eq
The recent CMB observation shows that $n_s=0.9649\pm0.0042 ~(68\% \mathrm{CL})$ \cite{Planck2018}. However, numerical results of the spectral index in the matter-Ekpyrotic bounce scenario studied in this paper turn out to be unfavored by the current observations. The details are depicted in Fig. \ref{f7a} where the initial conditions are set to the same as in (\ref{initial1}), in particular, $\mathcal E_\mathrm{dust}=2.00\times10^{-5}$. In the left panel of the figure, we show the plot for $p=0.1$ in the regime $k\in(10^{-5},2\times10^{-4})$, where the spectral index  larger than  unity  is observed for the modes therein. We find that this situation can not be improved by changing the parameters that characterize the Ekpyrotic potential. For example, in the right panel of the figure, we use a smaller $p$ which results in a narrower potential well.  The corresponding spectral index changes rapidly in the regime $k\gtrsim 5\times10^{-5}$ and remain almost constant ($\approx1$) in the regime $k\lesssim 5\times10^{-5}$. This is still inconsistent with the observations which requires a constant value of $n_s$ at approximately $0.9649$. In addition, we find that changing the dust energy density or the depth of the potential well does not produce a better prediction on the spectral index as well. Numerical analysis on $n_s$ with the initial conditions in (\ref{initial2}) gives similar results as depicted in Fig. \ref{f7a}.
One of the ways to address this issue is to have a quasi-matter dominated contracting phase (instead of an exact dust-dominated phase) with a slightly negative equation of state to produce a red tilt in the spectrum. Various ways of achieving this have been proposed, such as considering a $\Lambda$CDM bounce scenario \cite{CaiEwing2015}, or considering a specifically fine-tuned scalar field having a slightly negative equation of state \cite{WE1,haro2}, or considering dark matter and dark energy as the matter content \cite{CaiDuplessis2016,CaiEwing2016}.

\section{Conclusion} 
\label{conclusion}

In this paper, we have studied the background evolution of a spatially flat FLRW  universe in the matter-Ekpyrotic bounce scenario in the framework of LQC, obtained the numerical results of the primordial scalar power spectrum and discussed the spectral index predicted by the theory. The background evolution of the universe in this scenario is characterized by two distinct stages in the contracting phase: the matter-dominated phase and the scalar field dominated phase. The former is dominated by the dust field and thus endowed with a non-negative equation of state, while the latter is dominated by  a scalar field with a negative potential. In the scalar field dominated phase, there also exists the Ekpyrotic phase with an ultra-stiff equation of state which is necessary in general to avoid the BKL instability. The matter-dominated phase and the scalar field dominated  phase intersect at the transition point where the energy densities of the dust field and the Ekpyrotic field become equal to each other. We have used these transition points to compute the power spectra in contracting and expanding branches.

For the numerical analysis of the background dynamics, the initial conditions are imposed at the bounce where the parameter space is composed of the initial value of the Ekpyrotic field and the dust energy density.  With  the fixed  parameters for the Ekpyrotic potential and the same initial value of the Ekpyrotic field, we have varied  initial dust energy density, while keeping the Ekpyrotic field as the dominant component at the bounce, to investigate its effects on the duration of the scalar field dominated phase and the Ekpyrotic phase ($w>1$). A decrease in the fraction of the dust energy density at the bounce point can lead to a longer  duration of the scalar field dominated phase. In particular, as the dust energy density at the bounce decreases, the number of e-foldings for  the regime with $w>1$ increases. Besides, we find the width and the depth of the Ekpyrotic potential can also impact the duration of the regime with $w>1$ and the scalar  field dominated phase. In particular, increasing the width of the potential can prolong the duration of the $w>1$ regime as well as the scalar field dominated phase. On the other hand, increasing the depth of the potential reduces the scalar field dominated phase due to increased kinetic energy at the bottom of the well, while increasing the $w>1$ regime due to a deeper well.
In addition, the regime with $w>1$ before the bounce point only lasts a few number of e-foldings as the bounce is initially located near the bottom of the Ekpyrotic potential and dominated by the kinetic energy of the Ekpyrotic field. 

After studying background dynamics, we have applied the dressed metric approach in LQC to numerically analyze the propagation of the comoving curvature perturbations from the matter dominated contracting phase to the transition point in the expanding phase.  By comparing the power spectrum at different times during its evolution, we  found that the scale invariant regime of the power spectrum is already generated by the end of the matter dominated phase in the contracting phase. This regime lies in the range where the comoving wavenumbers are much less than unity. It turns out that in the matter-Ekpyrotic bounce scenario, the magnitude of the power spectrum in the scale invariant regime changes over time when the relevant modes propagate from the matter-dominated phase, across the bounce and then reach the transition point in the expanding regime. The varying magnitude of the power spectrum from our numerical analysis is  in contrast with the constant power spectrum obtained from the analytical approximations under the assumption of a constant Ekpyrotic equation of state \cite{CaiWE2014}. It is also different from the power spectrum in the inflationary scenario where magnitude is frozen after the relevant modes exit the Hubble horizon in the inflationary phase. Since the characteristic comoving wavenumber in the dressed metric approach of LQC is of the order of unity with bounce volume taken to unity,  the bouncing regime, including the Ekpyrotic phase, does not affect the qualitative behavior of the scale invariant regime.  The bounce only impacts  modes with the comoving wavenumbers around unity.  The magnitude of these affected modes  monotonically increase with the comoving wavenumber before the bounce and then  become oscillatory after the bounce.  Finally, the duration of the Ekpyrotic phase  not only directly  influences the magnitude of the scale invariant power spectrum observed at the transition point in the expanding phase but also changes the range of the comoving wavenumbers in the scale invariant regime of the power spectrum. 

From the numerical results of the scalar power spectrum, we have also  computed their corresponding  spectral index. Our results show the inconsistency between  the theoretical value of the spectral index in the current matter-Ekpyrotic bounce scenario and its experimental data. This inconsistency is due to the fact that the matter-dominated phase is sourced by the dust field which leads to an almost vanishing  equation of state. Therefore, the resulting spectral index in the scale invariant regime is very close to unity, which is excluded by the CMB observations.  We  have also found that changing the width and the depth of the Ekpyrotic potential or the dust energy density does not help improve the spectral index in the current scenario where the matter dominated phase is sourced by dust and the Ekpyrotic field in the potential given by (\ref{potential}).

Based on our studies, we think further exploration is required to establish a lower bound on the minimum number of e-foldings required to keep anisotropies from growing and preventing a BKL type instability in the context of LQC, which we aim to investigate in a future work using anisotropic spactimes.  Though this has been explored in case of a Galilean bounce \cite{CaiBBPP2013}, the results of such a study in LQC are expected to be different, given that in LQC anisotropic shear is bounded by quantum geometry \cite{pswe}. Further issues for exploration include the tensor-to-scalar ratio, constraints on non-Gaussianities, and a graceful exit strategy for the matter bounce scenario to the reheating phase, leading to our current observational universe. Several proposals exist for addressing each one of these issues for the matter bounce scenario, but it is non-trivial to satisfy the observational constraints on all these aspects at the same time for a given model especially with inclusion of an Ekpyrotic phase. It will be also interesting to explore the consequences for the matter bounce family of models incorporating more loop quantum gravity effects such as gauge-covariant fluxes recently studied in \cite{liegener-singh} which can bring non-trivial changes to dynamics and resulting predictions. Finally, a limitation of the current analysis is to rely on single fluid treatment in perturbations and it will be important to generalize to the setting of two-fluid model for perturbations in LQC effective dynamics.

In summary, we have numerically analyzed a specific realization of the matter-Ekpyrotic bounce scenario by considering dust and an Ekpyrotic field in a negative potential as the matter content in the effective dynamics of LQC. Our numerical results show that the existence of the Ekpyrotic phase and, namely a regime with $w>1$ right before the bounce is robust with respect to the variations of the parameters in the Ekpyrotic potential and the initial conditions. This is also the first time  that the dressed metric approach is applied to compute the power spectrum in the matter bounce scenario in LQC. From this approach, we find  a scale-invariant regime and an oscillatory regime in the power spectrum. Although further studies on the spectral index reveal inconsistency between its predicted value and the observational data which could potentially be alleviated by considering a quasi-matter dominated contracting phase with a slightly negative equation of state.

\section*{Acknowledgements} 
This work is supported by the NSF grants PHY-1454832.

\end{document}